\documentclass[preprint,superscriptaddress]{revtex4}
\usepackage{graphicx}

\newcommand{\bra}[1] {\left\langle #1 \right|}
\newcommand{\ket}[1] {\left| #1 \right\rangle}

\begin{document}

\title{Properties and dynamics of generalized squeezed states}

\author{Sahel Ashhab}
\affiliation{Advanced ICT Research Institute, National Institute of Information and Communications Technology (NICT), 4-2-1, Nukui-Kitamachi, Koganei, Tokyo 184-8795, Japan}

\author{Mohammad Ayyash}
\affiliation{Institute for Quantum Computing, University of Waterloo, 200 University Avenue West, Waterloo, Ontario N2L 3G1, Canada}
\affiliation{Red Blue Quantum Inc., 72 Ellis Crescent North, Waterloo, Ontario N2J 3N8, Canada}

\begin{abstract}
We analyze the properties and dynamics of generalized squeezed states. We find that, in stark contrast to displacement and two-photon squeezing, higher-order squeezing leads to oscillatory dynamics. The state is squeezed in the initial stages of the dynamics but the squeezing reverses at later stages, and the state reverts almost completely back to the initial state. We analyze various quantities to verify that the oscillatory dynamics is physical and not a mathematical artefact. We also show that the maximum squeezing diminishes with increasing squeezing order, rendering the squeezing mechanism increasingly ineffective. Our results provide important rules that can help guide the development of more effective higher-order squeezing techniques.
\end{abstract}

\maketitle

\newpage

\section{Introduction}
\label{Sec:Introduction}

Remarkable advances have been made in the past few decades in the field of quantum information technology \cite{Nielsen,Fedorov,Kimble,Wehner,Illiano,Degen}. Quantum computers with hundreds of qubits are already in commercial use. Quantum secure communication systems are being constructed around the world. Quantum sensing techniques have achieved precision that surpasses classical sensing limits. More progress on all these technologies is expected in the coming years.

One of the important resources in quantum technologies are photonic states \cite{Walls,Scully}. Recent years have seen the development and demonstration of new capabilities in controlling the states of light, including the generation of nonclassical states, which are crucial for quantum communication applications. The generation of nonclassical states requires the use of some nonlinear mechanism, since linear optics cannot turn classical states of light into nonclassical ones \cite{Drummond}.

For many years, the study of nonlinear effects in light relied on combining strong input lasers with weak quadratic nonlinearities to generate weak but nonclassical output photon states. In recent years, especially with the development of circuit quantum electrodynamics in superconducting circuits, it has become possible to achieve much stronger nonlinearities and demonstrate various nonlinearity-based effects \cite{Yurke,Movshovich,Yamamoto,Kudra}. Implementations of nonlinear phenomena at to the single photon level are becoming increasingly common. Furthermore, it is now possible to access higher-order nonlinearities, e.g.~cubic nonlinearities \cite{Chang,Eriksson,Bazavan,Saner}.

One of the most extensively studied types of nonclassical states are squeezed states. Two-photon squeezed states, which can be obtained by using quadratic nonlinearities, have been investigated for decades. Although these states are nonclassical, they are still Gaussian, which limits their usefulness for quantum information processing. It is therefore important to investigate higher-order squeezed states, which are non-Gaussian and promise new applications in quantum computation and communication \cite{Ghose,Hillery1999,DellAnno,Zheng}. For example, higher-order squeezing could be used to generate multi-photon entangled states \cite{DellAnno}. Furthermore, combining a non-Gaussian operation with displacement and squeezing can be a way to achieve universal control of photonic states \cite{Eriksson}. In the past few years, there have been a few demonstrations of higher-order squeezed states \cite{Chang,Eriksson,Bazavan,Saner}, as well as theoretical proposals for utilizing higher-order nonlinearities in physical setups \cite{Koshino,Ayyash}. While there have been some theoretical studies on this subject \cite{Fisher,Hillery1984,Hong,Braunstein1987,Braunstein1990,Hillery1990,Zelaya,Braak}, as well as the related subject of three-photon down-conversion \cite{Banaszek,Felbinger,Olsen,Bencheikh}, a full picture of higher-order squeezing is lacking.

In this work we aim to add insight to this area that is expected to play a role in future nonlinear quantum optics applications. We therefore perform a systematic analysis of the properties and dynamics of generalized squeezed states. We find some unexpected results, especially relating to periodic dynamics, that differ qualitatively from what is known about the lower-order operations of displacement and two-photon squeezing. As the squeezing order is increased, the deviation of the quantum state from the initial state decreases and the quantum state returns to the initial state more rapidly. In other words, the maximum amount of squeezing that can be achieved decreases with increasing squeezing order.

The remainder of this paper is organized as follows: in Sec.~\ref{Sec:Background}, we provide some background information about previous work, and we describe our numerical simulation methodology. In Sec.~\ref{Sec:Results}, we present the results of our numerical simulations, including sequences of Husimi and Wigner functions, as well as plots depicting the dependence of the photon number on the squeezing parameter. We discuss the interpretation of our numerical simulation results and their implications for physical realizations of generalized squeezing. In Sec.~\ref{Sec:Discussion}, we discuss the results from a few additional points of view. In Sec.~\ref{Sec:Conclusion}, we end with some concluding remarks.

\section{Background and methodology}
\label{Sec:Background}

Two well-known operations in quantum optics are displacement and two-photon squeezing. The former is described by the unitary operator
\begin{equation}
\hat{D} \left( \alpha \right) = \exp \left\{ \alpha \hat{a}^\dagger - \alpha^* \hat{a} \right\},
\label{Eq:DisplacementOp}
\end{equation}
while the latter is described by the unitary operator
\begin{equation}
\hat{S} \left( \xi \right) = \exp \left\{ \left[ \xi \left(\hat{a}^\dagger\right)^2 - \xi^* \hat{a}^2 \right] / 2 \right\},
\label{Eq:SqueezingOp}
\end{equation}
where $\hat{a}$ and $\hat{a}^{\dagger}$ are, respectively, the photon annihilation and creation operators. These operations have been extensively analyzed in the literature. It is worth emphasizing that these two cases allow analytical treatment. For example, there are well-known exact solutions for the final state when one of these operators is applied to the vacuum state $\ket{0}$, namely coherent states for the former and squeezed vacuum states for the latter. Furthermore, there are relations that allow straightforward prediction of what happens when a sequence of these operations are combined and applied to an input quantum state.

It is then natural to consider the $n$th-order (or $n$-photon) generalization of displacement and two-photon squeezing:
\begin{equation}
\hat{U}_n \left( r \right) = \exp \left\{ r \left(\hat{a}^\dagger\right)^n - r^* \hat{a}^n \right\}.
\label{Eq:GeneralizedSqueezingOp}
\end{equation}
To simplify the discussion, we shall restrict our analysis to real values of $r$. We note, however, that the squeezing operator for any complex value of $r$ can be obtained by combining the squeezing operator for $|r|$ with a simple rotation \cite{Braunstein1987}. We note also that, as is commonly done in the literature on generalized squeezing, we do not include any extra coefficient in Eq.~(\ref{Eq:GeneralizedSqueezingOp}), in contrast to the factor of 2 that appears at the end of Eq.~(\ref{Eq:SqueezingOp}).

It is interesting to start the discussion by commenting on a theoretical complication that was discussed in the early literature on generalized squeezing. Fisher {\it et al.} \cite{Fisher} pointed out that the power series expansion
\begin{equation}
\bra{0} \exp \left\{ r \left[ \left(\hat{a}^\dagger\right)^n - \hat{a}^n \right] \right\}  \ket{0} = 1 + r \bra{0} \left[ \left(\hat{a}^\dagger\right)^n - \hat{a}^n \right]  \ket{0} + \frac{r^2}{2} \bra{0} \left[ \left(\hat{a}^\dagger\right)^n - \hat{a}^n \right]^2  \ket{0} + ...
\label{Eq:PropagatorExpansion}
\end{equation}
leads to divergences. Specifically, they showed that the absolute values of the individual terms in the series grow indefinitely. They concluded that the squeezing operation is unphysical. It was later shown that this pathological behaviour, which occurs for $n\geq 3$, pertains only to the power series expansion of the exponential function and that a more careful analysis of the squeezing dynamics can avoid this mathematical artefact \cite{Braunstein1987,Braunstein1990}.

In fact, one can express the generalized squeezing operator as
\begin{equation}
\hat{U}_n \left( r \right) = \exp \left\{ r \left[ \left(\hat{a}^\dagger\right)^n - \hat{a}^n \right] \right\} = \exp \left\{ -i \hat{H}_n r \right\},
\label{Eq:Propagator_n}
\end{equation}
where the generator
\begin{equation}
\hat{H}_n = i \left[ \left(\hat{a}^\dagger\right)^n - \hat{a}^n \right]
\label{Eq:Hamiltonian_n}
\end{equation}
is a Hermitian operator. One can then use physics intuition to think about the squeezing operator by treating the squeezing parameter $r$ as the dimensionless time variable and treating $\hat{H}_n$ as the effective dimensionless Hamiltonian that generates the squeezing dynamics. From this point of view, it is straightforward to conclude that the operator $\hat{U}_n \left( r \right)$ is unitary and therefore cannot have any divergent matrix elements.

In this work, we use numerical calculations to investigate the effect of the generalized squeezing operator on the vacuum state. We perform a variety of checks to look for any indications that unphysical mathematical artefacts might be appearing in the numerical results. As we shall explain below, we do not see any such indications.

Our calculations proceed as follows: we construct a large $N \times N$ matrix that represents a truncated version of the Hamiltonian in Eq.~(\ref{Eq:Hamiltonian_n}). Since all the matrix elements in $\hat{H}_n$ are equal to zero except between states that are different by exactly $n$ photons, we can divide the $N \times N$ matrix into $n$ smaller matrices of size $(N/n) \times (N/n)$, assuming that $N/n$ is an integer. Each one of these $n$ matrices corresponds to one subspace of the full Hilbert space. The $n$ orthogonal subspaces are $\left\{ \ket{0},\ket{n},\ket{2n},...\right\}$, $\left\{\ket{1}, \ket{n+1}, \ket{2n+1},...\right\}$, ..., $\left\{ \ket{n-1}, \ket{2n -1}, \ket{3n -1},... \right\}$. For example, for $n=3$, we have three subspaces that do not mix: $\left\{\ket{0}, \ket{3}, \ket{6}, ...\right\}$, $\left\{\ket{1}, \ket{4}, \ket{7}, ...\right\}$ and $\left\{\ket{2}, \ket{5}, \ket{8}, ...\right\}$. This separation of the full $N \times N$ matrix into $n$ smaller ones simplifies further manipulation of the matrices. Once we have constructed one of the relevant matrices that represent $\hat{H}_n$, we evaluate the corresponding squeezing operator for a small value of the squeezing parameter $\delta r$ by taking the exponential of the truncated Hamiltonian matrix multiplied by $-i\delta r$, i.e.~we evaluate $\hat{U}_n (\delta r)=\exp\left\{-i\delta r \hat{H}_n\right\}$. We always use $\delta r =0.01$. We evaluate the squeezing operator for larger values of $r$ by multiplying $\hat{U}_n (\delta r)$ by itself $r/\delta r$ times. In other words, $\hat{U}_n (r)=\left[\hat{U}_n (\delta r)\right]^{r/\delta r}$. We performed a few random checks by evaluating $\hat{U}_n (r)$ in one step as $\exp\left\{-i r \hat{H}_n\right\}$ and inspecting the amplitudes in the state $\hat{U}_n (r)\ket{0}$. We found that the results differed from those obtained when using the formula $\hat{U}_n (r)=\left[\hat{U}_n (\delta r)\right]^{r/\delta r}$ by only $\sim 10^{-10}$, which is an indication that we are not encountering any serious issues with the numerical evaluation of $\exp\left\{-i r \hat{H}_n\right\}$. Since some previous studies that performed numerical simulations of generalized squeezing had the issue of an unconserved total probability, we point out that in our simulations the quantum state always remained normalized, as it should. We vary the squeezing parameter $r$, evaluate $\hat{U}_n (r)\ket{0}$ and analyze the state as a functions of $r$.

\begin{figure}[h]
\includegraphics[width=9cm]{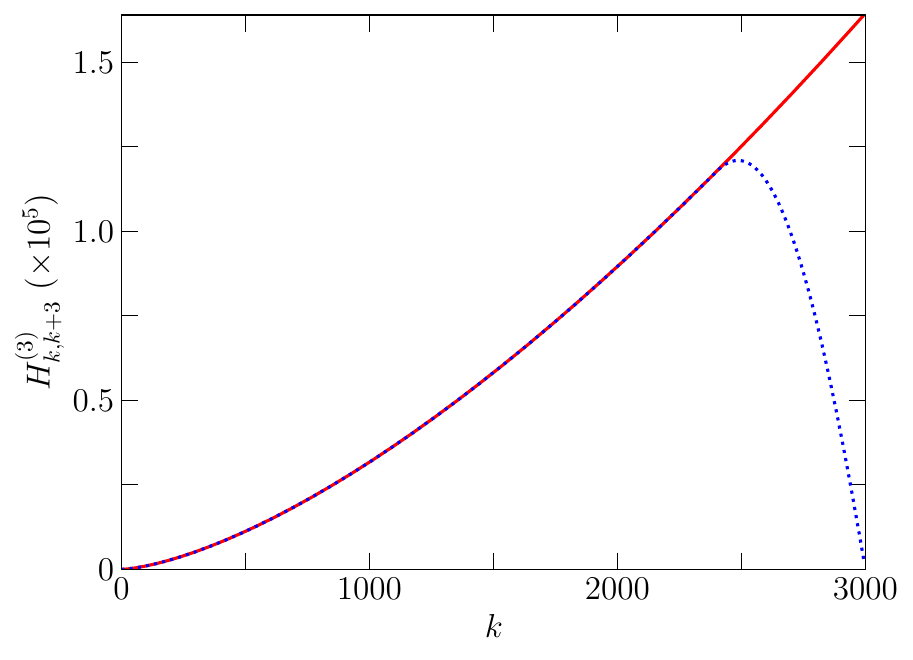}
\caption{Matrix elements for the matrix $\hat{H}_3$ truncated by keeping a total of $N=3000$ photon number states. The red solid line and blue dotted line correspond, respectively, to the cases of hard and soft cutoff.}
\label{Fig:MatrixElements}
\end{figure}

As mentioned above, one must take care to avoid divergences that arise in the power series expansion of the exponential function $\exp\left\{-i\delta r \hat{H}_n\right\}$ when $n\geq 3$. We used the built-in matrix exponentiation functions in the software packages Mathematica (Version 13.2.1.0) and Numpy (Version 1.19.2). Both consistently showed no evidence of divergences, and they agreed with each other to the level of numerical precision ($\sim 10^{-14}$).

We also note that the matrix elements of $\hat{H}_n$ are defined by the main rule
\begin{equation}
H_{k,k+n}^{(n)} = -i \sqrt{(k+1) (k+2) \cdots (k+n)},
\label{Eq:MatrixElements}
\end{equation}
where we have moved the index $n$ to the superscript to avoid having a subscript with three indices. An important factor in giving rise to divergences is the fact that, for large $n$, the matrix elements grow very fast for large $k$. As a result, we must be careful and try to analyze how our results depend on the truncated matrix size, so that we can make predictions about the results for an infinite, untruncated matrix. For this purpose, we perform several calculations with the same system parameters but with different matrix sizes, and we investigate how our results change as we vary the matrix size $N$. As a further check on the effect of matrix truncation on our results, we construct soft-cutoff matrices in which we take the original (hard-cutoff) $N \times N$ Hamiltonian matrix and multiply the matrix elements in the bottom-right corner of the matrix by a suppression factor, such that the matrix elements in the bottom-right corner of the matrix decrease and approach zero smoothly. The specific suppression factor that we use is equal to 1 for $k/N<0.8$, such that all matrix elements with $k<0.8 N$ remain unchanged. However, when $k/N>0.8$, we multiply the matrix element $H_{k,k+n}^{(n)}$ by the factor $\sin[5\pi (N-k)/(2N)]$, which decreases from 1 at $k=0.8 N$ to zero at $k=N$. The matrix element dependence on $k$ for the cases of hard and soft cutoff are illustrated in Fig.~\ref{Fig:MatrixElements}. In the case of the hard cutoff, the matrix elements keep increasing all the way up to the highest value of $k$ kept in the truncated matrix. This situation is analogous to having a hard wall when studying the dynamics of a moving particle. In the case of the soft cutoff, the matrix elements decrease smoothly as we approach the high-$k$ edge of the truncated matrix. We note that the details of our soft-cutoff suppression factor are arbitrary. We expect that our conclusions will remain the same for any generally similar soft-cutoff truncation. In fact, we performed a few test simulations in which the soft cutoff was introduced by adding gradually increasing diagonal matrix elements in the bottom-right corner of $\hat{H}_n$. The results of these simulations were qualitatively similar to the results with the other cutoff conditions that we used in this study.

Our approach of comparing different matrices that all approximate $\hat{H}_n$ can also be interpreted differently from just being a check on whether our numerical approach is sound or not. From a practical point of view, one can say that, while the behaviour of an infinite perfect system is of pure theoretical interest, it is important to identify results that are robust to variations in the Hamiltonian away from the perfect form given in Eq.~(\ref{Eq:Hamiltonian_n}). In a practical situation, e.g.~in a superconducting circuit containing nonlinear elements whose purpose is to induce the generalized squeezing effect, we do not expect that the physical system will be accurately described by the Hamiltonian in Eq.~(\ref{Eq:Hamiltonian_n}) beyond tens of photons. A system that contains a cubic nonlinearity will almost invariably contain quartic and higher-order nonlinearities whose effects become increasingly nonnegligible with increasing photon number. These effects can naturally impose a cutoff on the number of photons up to which Eq.~(\ref{Eq:Hamiltonian_n}) is a good approximation. From this point of view, it might make more sense to use the finite-$N$ simulation results as a reference point to test and judge the soundness and/or relevance of infinite-$N$ treatments, rather than the other way around.

It is also worth mentioning here that the Hamiltonian in Eq.~(\ref{Eq:Hamiltonian_n}) is unbounded from below, i.e.~it predicts infinite negative energies, and can therefore be considered unphysical. One must note, however, that this Hamiltonian is typically obtained as an effective Hamiltonian in a rotating frame. If the initial state has a small number of photons, the dynamics generated by the Hamiltonian will not take the system to the infinite-photon-number region, where the unphysical effects are encountered. It should be noted that the issue of boundedness applies also for the cases $n=1$ and $n=2$, where it is known that this Hamiltonian, or truncated versions of it, give accurate descriptions of the dynamics.

\section{Results}
\label{Sec:Results}

\begin{figure}[h]
\includegraphics[width=4cm]{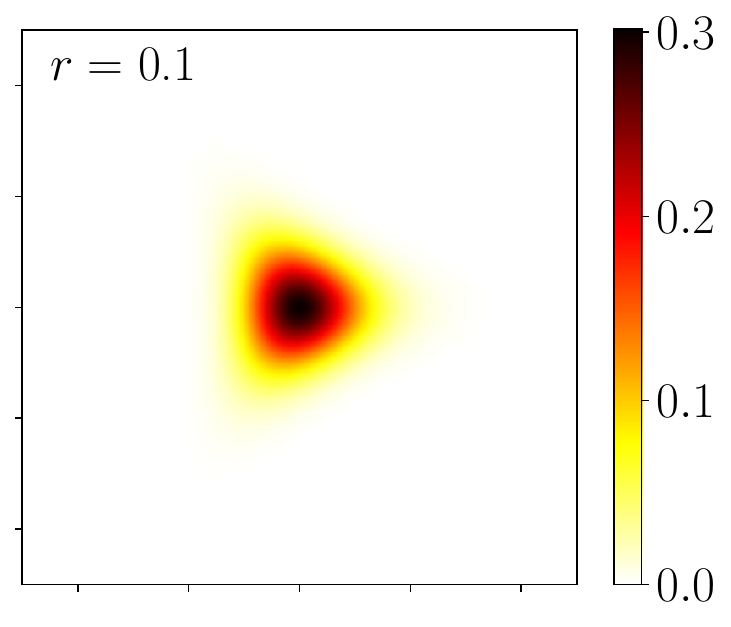}
\includegraphics[width=4cm]{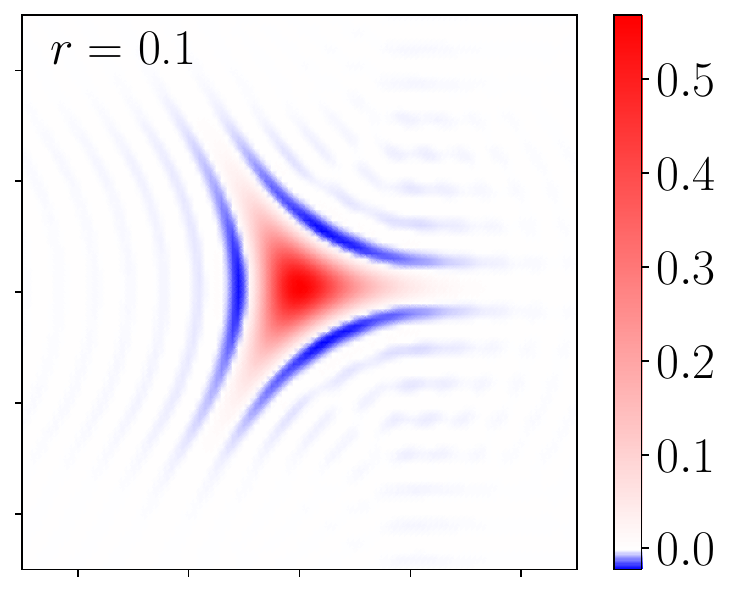}
\includegraphics[width=4cm]{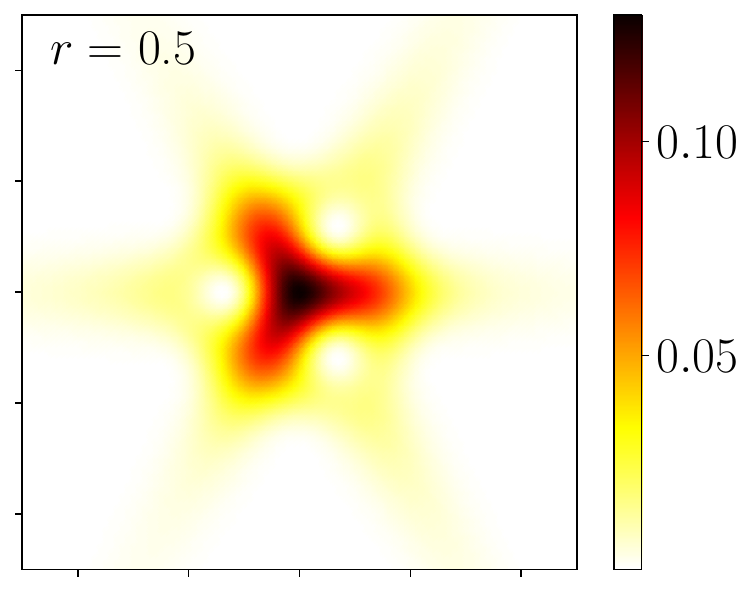}
\includegraphics[width=4cm]{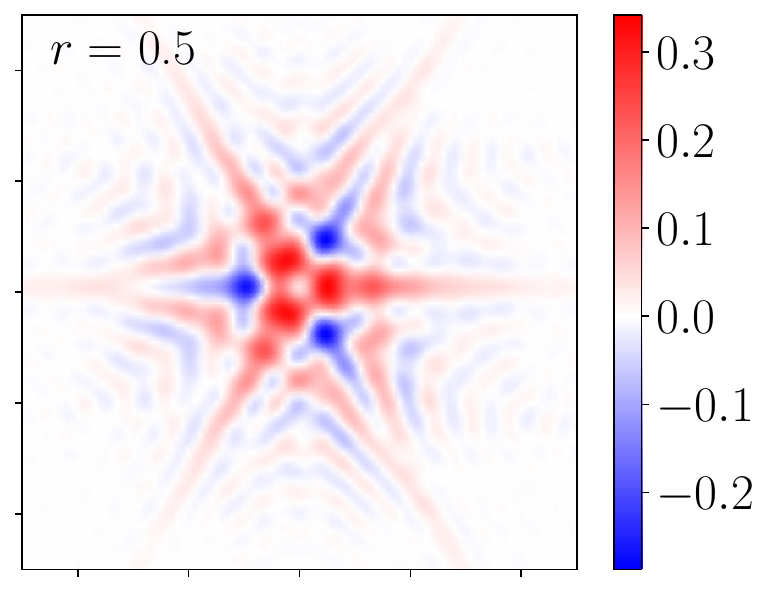}
\includegraphics[width=4cm]{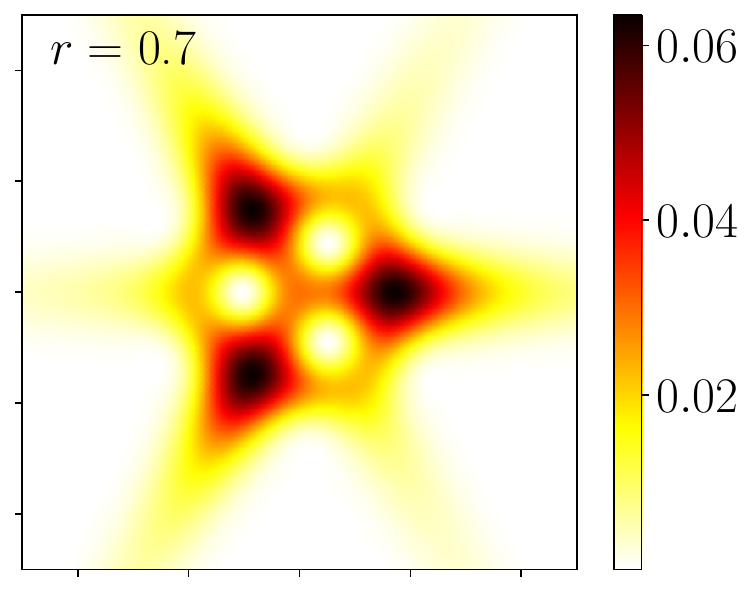}
\includegraphics[width=4cm]{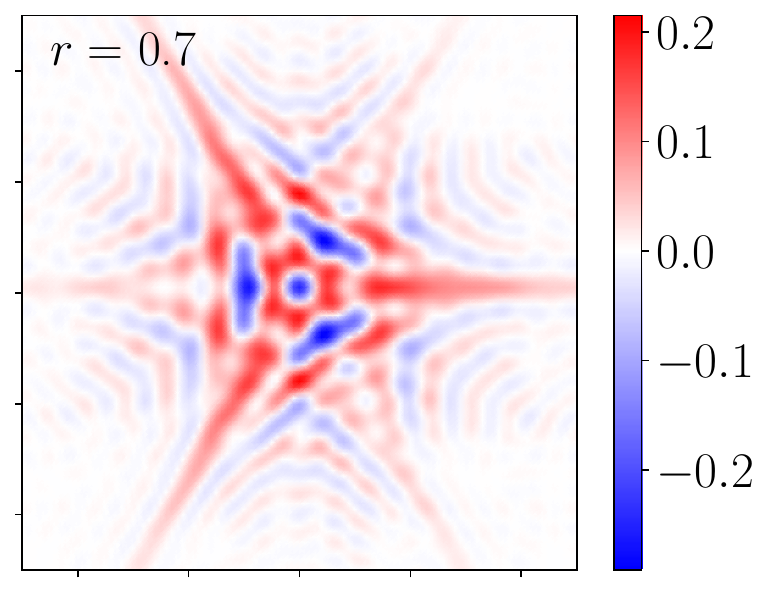}
\includegraphics[width=4cm]{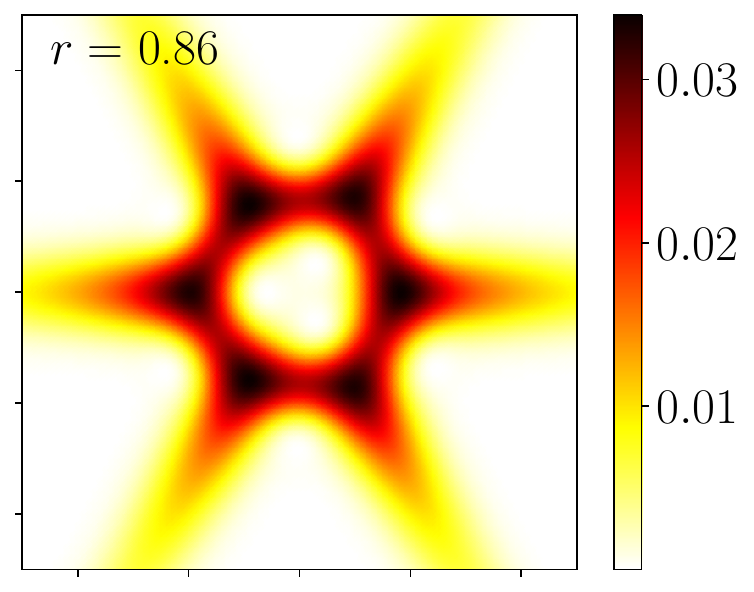}
\includegraphics[width=4cm]{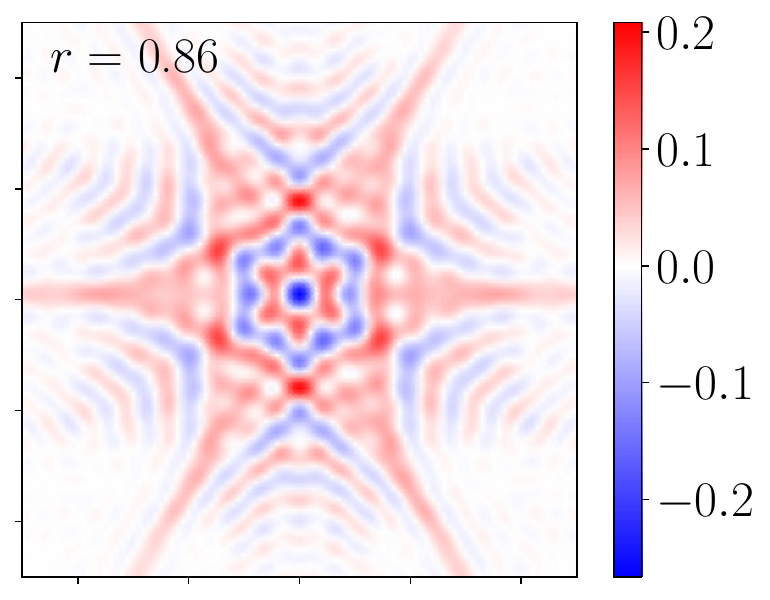}
\includegraphics[width=4cm]{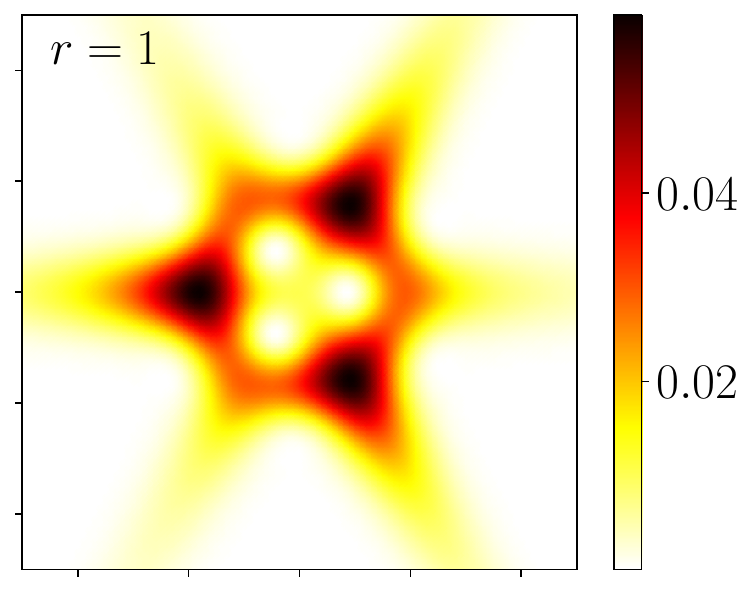}
\includegraphics[width=4cm]{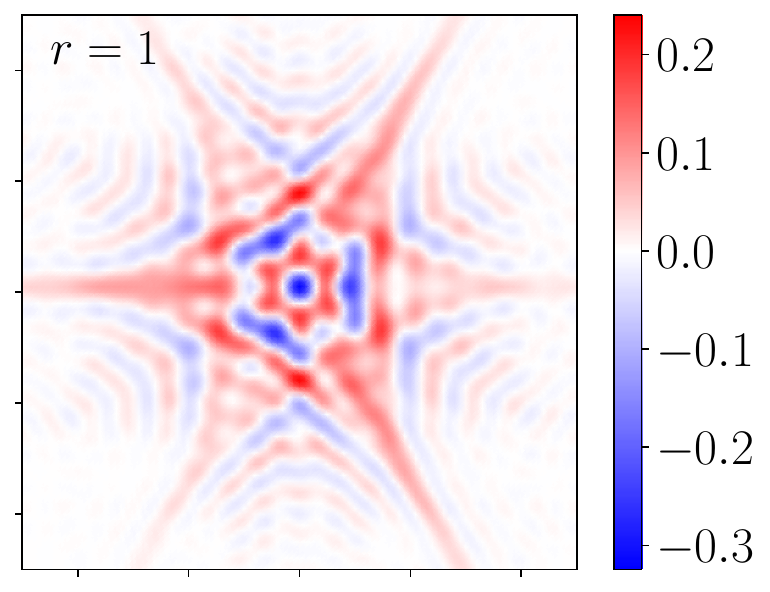}
\includegraphics[width=4cm]{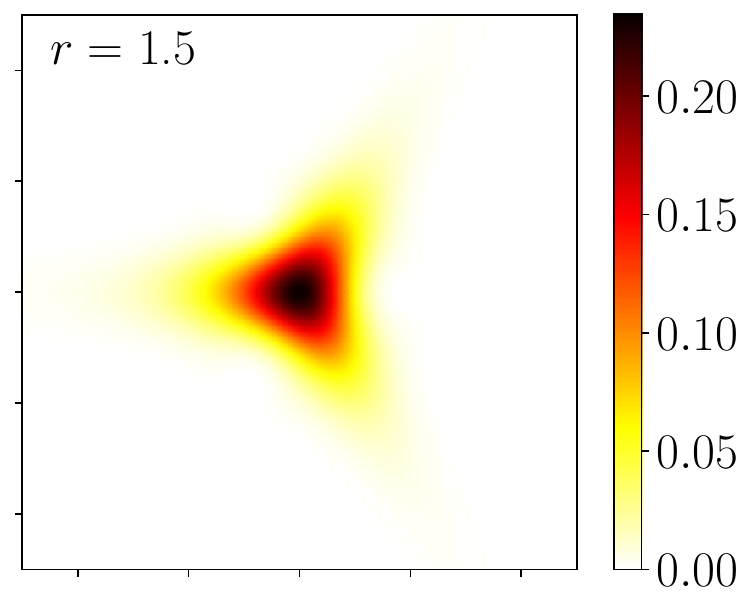}
\includegraphics[width=4cm]{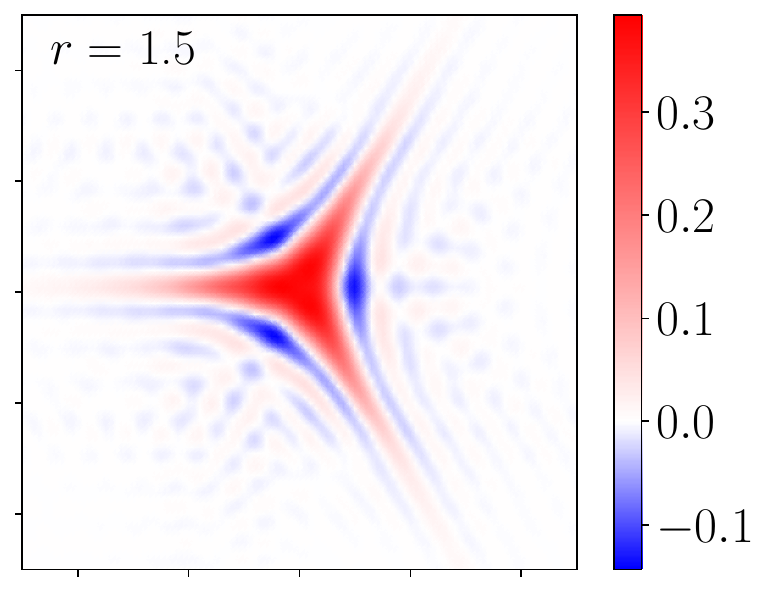}
\caption{Q and Wigner functions for the trisqueezed vacuum state, i.e.~$\hat{U}_3(r)\ket{0}$, for $r=0.1$, 0.5, 0.7, 0.86, 1 and 1.5. In all Q and Wigner function plots in this manuscript, the $x$ and $y$ axes correspond, respectively, to the position ($x$) and momentum ($p$) variables of phase space. The range of the axes is $[-5,5]$.}
\label{Fig:QW3}
\end{figure}

\begin{figure}[h]
\includegraphics[width=9cm]{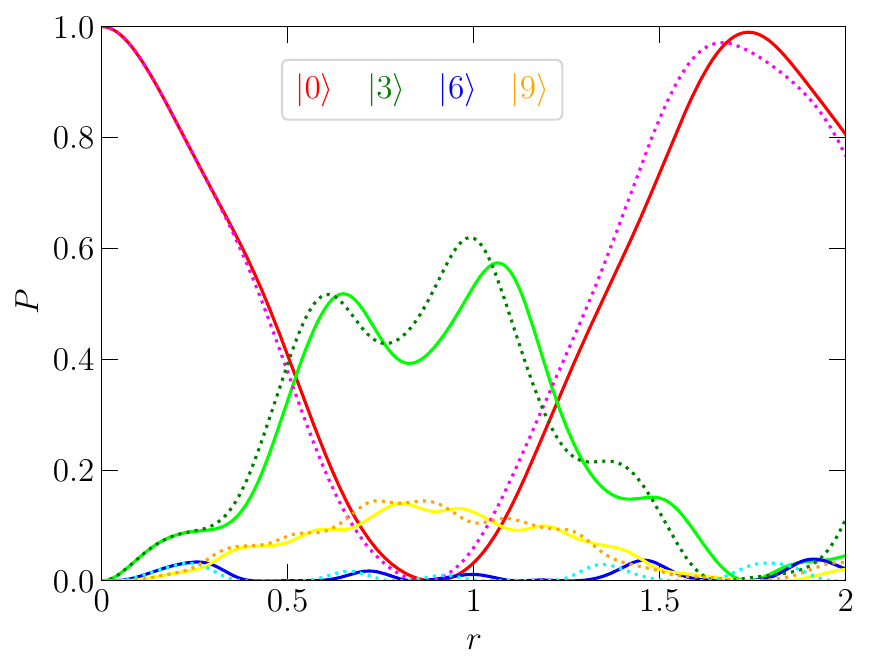}
\caption{Occupation probabilities of the lowest four relevant Fock states (i.e.~$\ket{0}$, $\ket{3}$, $\ket{6}$ and $\ket{9}$) for the trisqueezed vacuum state $\hat{U}_3(r)\ket{0}$ as functions of the squeezing parameter $r$. The solid lines correspond to a hard-cutoff approximation of $\hat{H}_n$ with $N=3\times 10^4$, while the dotted lines correspond to a soft cutoff with $N=300$. The red/magenta, light/dark green, blue/cyan and yellow/orange lines correspond, respectively, to the states $\ket{0}$, $\ket{3}$, $\ket{6}$ and $\ket{9}$. Despite the vast difference in the size of the truncated matrices, the difference in the probabilities is relatively small. Both data sets exhibit almost periodic oscillations with a period of approximately $r=1.7$.}
\label{Fig:ProbVsSqueezeingR3}
\end{figure}

\begin{figure}[h]
\includegraphics[width=9cm]{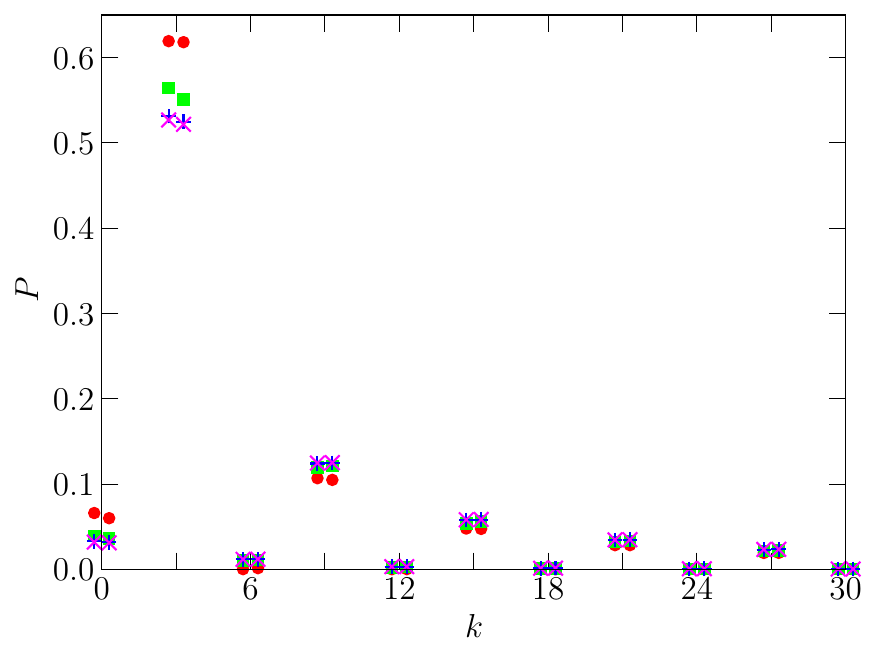}
\caption{Occupation probabilities of the lowest 11 relevant states ($\ket{0}$, $\ket{3}$, $\ket{6}$, ... $\ket{30}$) for the trisqueezed vacuum state with $r=1$, i.e.~$\hat{U}_3(r=1)\ket{0}$, as functions of photon number $k$. Each value of $k$ has eight different data points. The ones shifted slightly to the left correspond to a hard cutoff, while the ones shifted slightly to the right correspond to a soft cutoff. The red dots, green squares, blue + signs and magenta $\times$ symbols correspond, respectively to $N=300$, 3000, $3\times 10^4$ and $6\times 10^4$. The results are almost independent of the truncation details except for two points, namely the states $\ket{0}$ and $\ket{3}$. We performed additional simulations with $N=1.2 \times 10^4$, $N=2.1 \times 10^4$ and $N=4.5 \times 10^4$, and the results (not shown) were very close to those with $N=3 \times 10^4$ and $N=6 \times 10^4$.}
\label{Fig:ProbVsPhotonNum3}
\end{figure}

\begin{figure}[h]
\includegraphics[width=4cm]{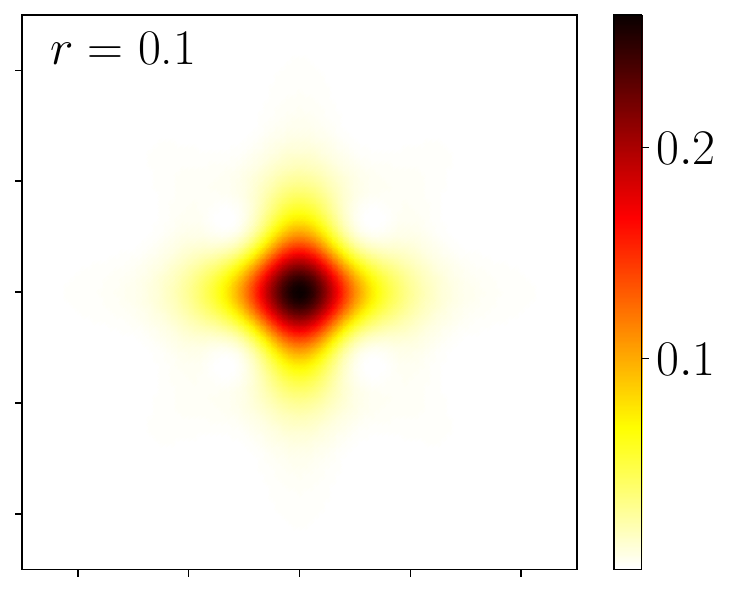}
\includegraphics[width=4cm]{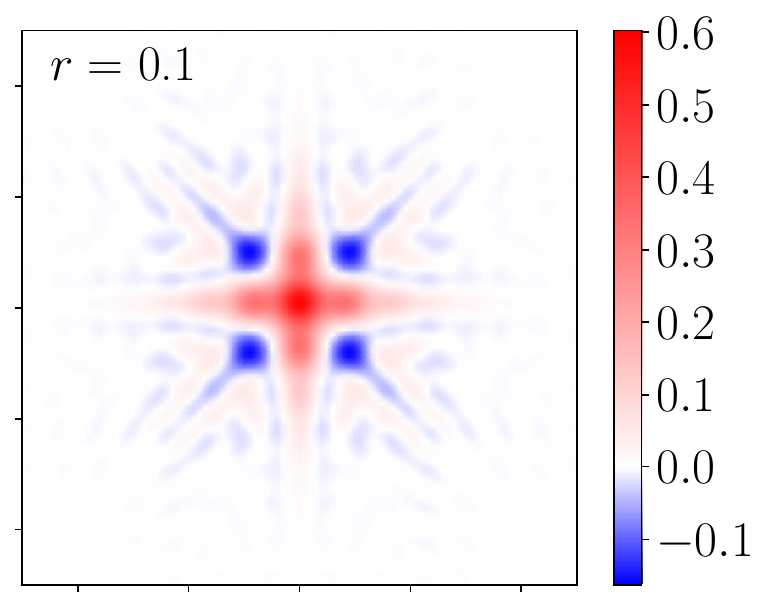}
\includegraphics[width=4cm]{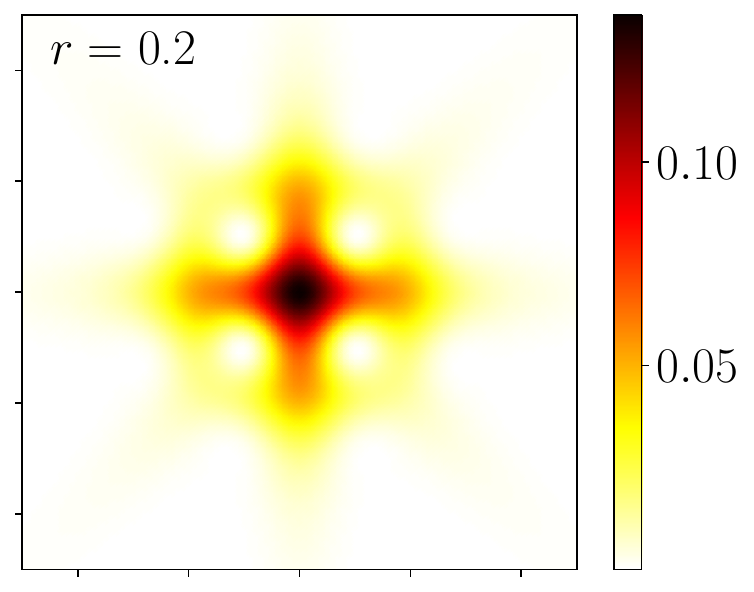}
\includegraphics[width=4cm]{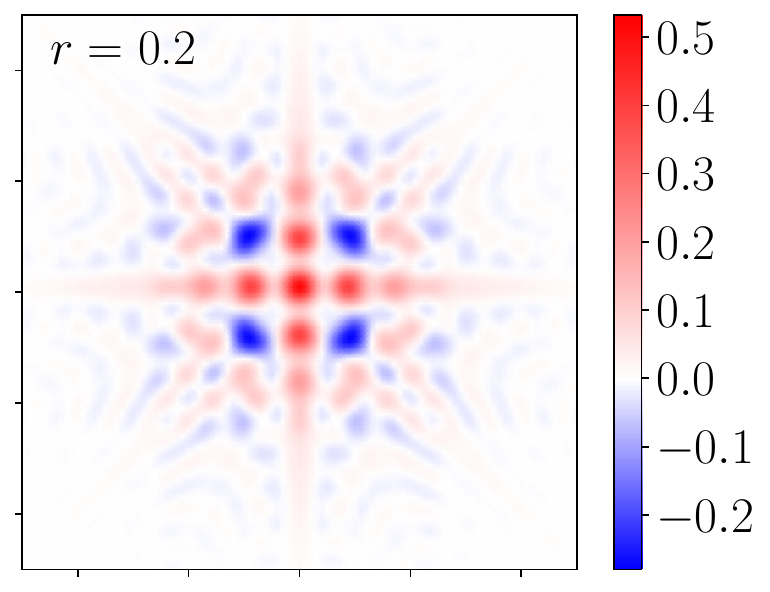}
\includegraphics[width=4cm]{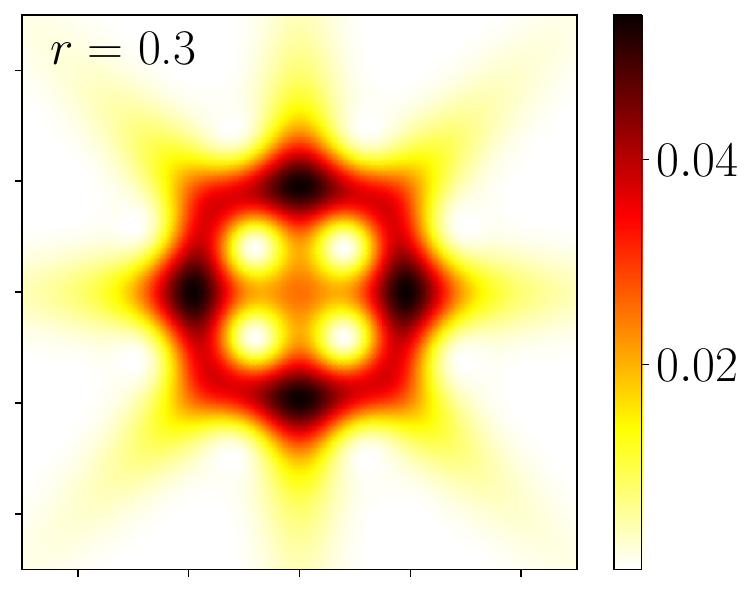}
\includegraphics[width=4cm]{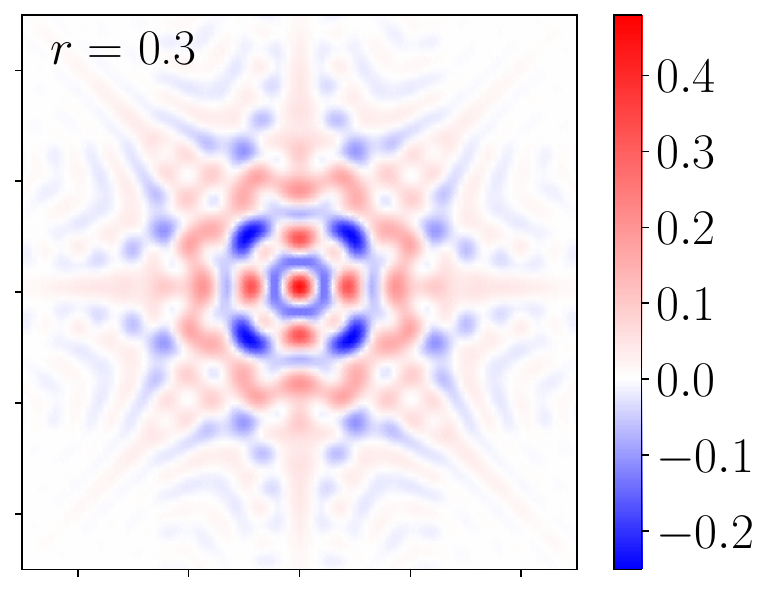}
\includegraphics[width=4cm]{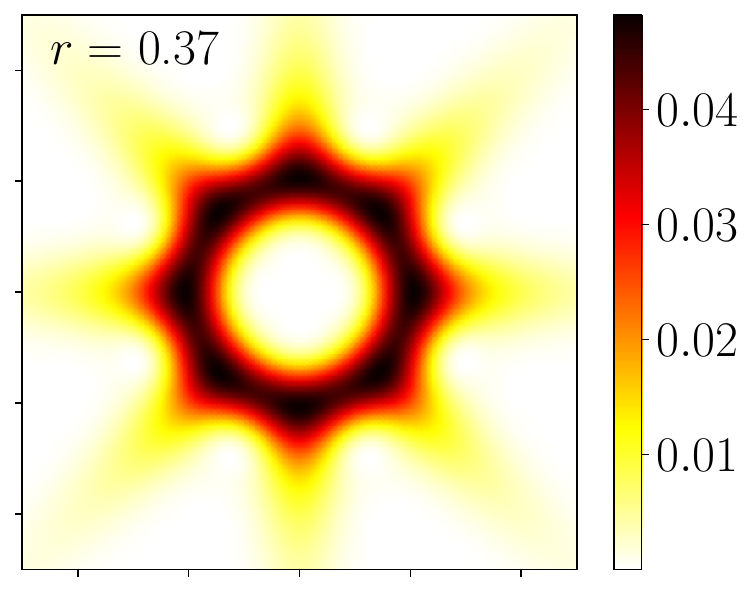}
\includegraphics[width=4cm]{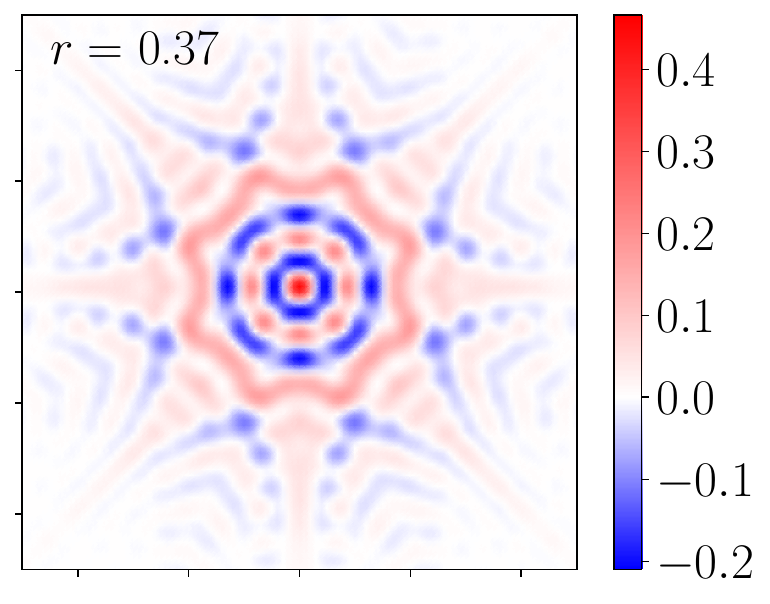}
\includegraphics[width=4cm]{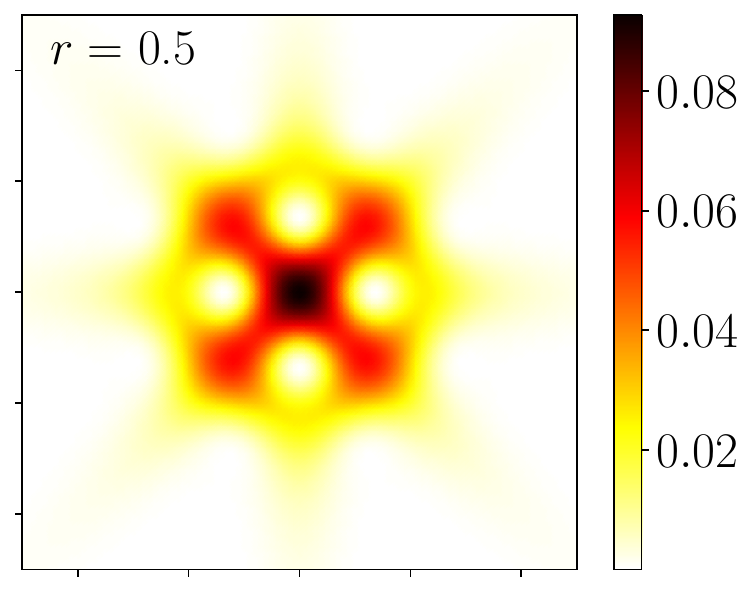}
\includegraphics[width=4cm]{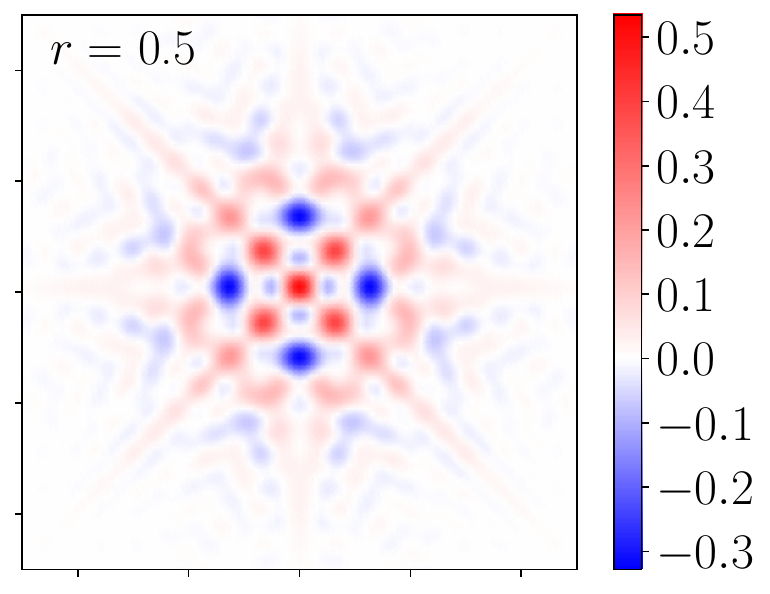}
\includegraphics[width=4cm]{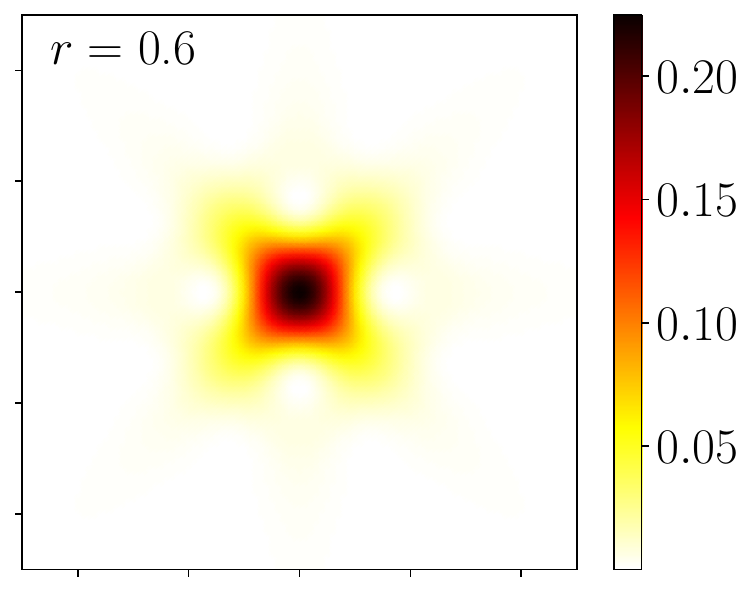}
\includegraphics[width=4cm]{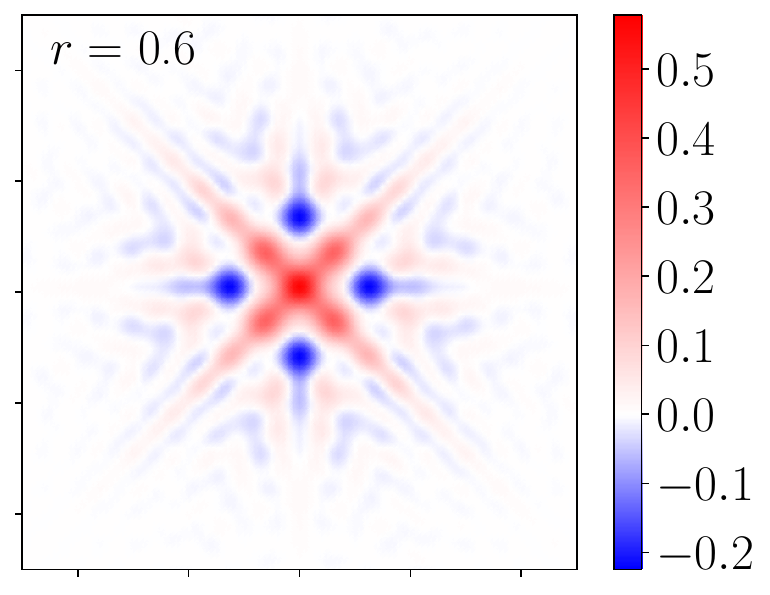}
\caption{Q and Wigner functions for the four-photon squeezed vacuum state, i.e.~$\hat{U}_4(r)\ket{0}$, for $r=0.1$, 0.2, 0.3, 0.37, 0.5 and 0.6.}
\label{Fig:QW4}
\end{figure}

\begin{figure}[h]
\includegraphics[width=9cm]{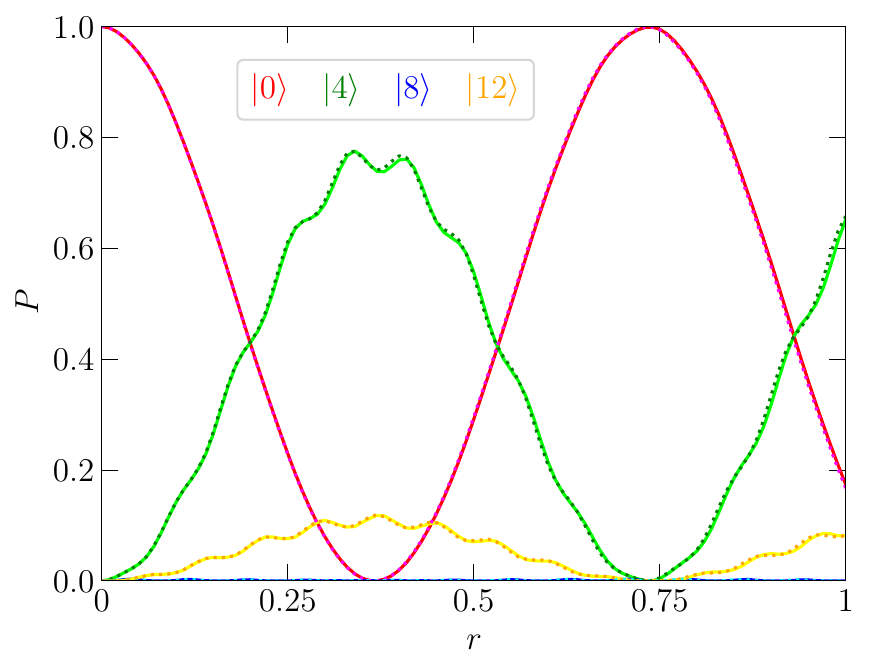}
\caption{Occupation probabilities of the lowest four relevant Fock states (i.e.~$\ket{0}$, $\ket{4}$, $\ket{8}$ and $\ket{12}$) for the four-photon squeezed vacuum state $\hat{U}_4(r)\ket{0}$ as functions of the squeezing parameter $r$. The solid lines correspond to a hard cutoff with $N=4\times 10^4$, while the dotted lines correspond to a soft cutoff with $N=400$. The red/magenta, light/dark green, blue/cyan and yellow/orange lines correspond, respectively, to the states $\ket{0}$, $\ket{4}$, $\ket{8}$ and $\ket{12}$. The two data sets are essentially indistinguishable.}
\label{Fig:ProbVsSqueezeingR4}
\end{figure}

\begin{figure}[h]
\includegraphics[width=8cm]{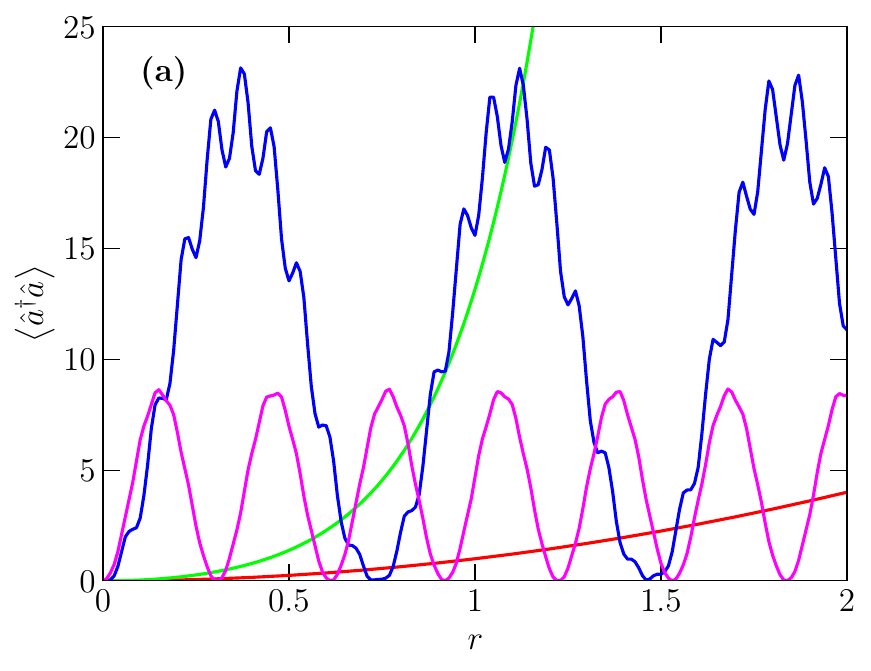}
\includegraphics[width=8cm]{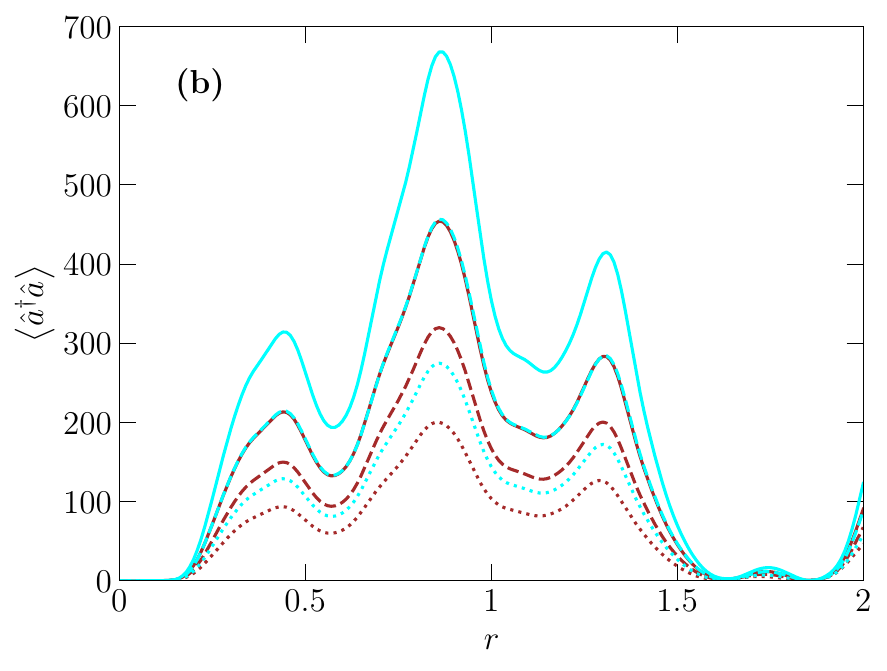}
\caption{Average photon number $\left\langle \hat{a}^{\dagger} \hat{a} \right\rangle$ for the state $\hat{U}_n(r)\ket{0}$ as a function of the squeezing parameter $r$. In Panel (a), the red, green, blue and magenta lines correspond, respectively, to $n=1$, 2, 4 and 5, all calculated with $N=n\times 10^4$ and a hard cutoff. The cases $n=1$ and $n=2$ correspond, respectively, to displacement and two-photon squeezing. We plot the results for $n=3$ separately (Panel b), because it requires a much higher $y$-axis scale. The brown and cyan lines correspond, respectively, to hard and soft cutoffs. The solid, dashed and dotted lines correspond, respectively, to $N=1.2\times 10^4$, $3\times 10^4$ and $6\times 10^4$. While the vertical scale of the different lines depends on the Hamiltonian truncation details, the overall oscillation pattern seems to be almost completely independent of these details. This point is most clearly seen in the coincidental overlap between the solid brown and dashed cyan lines.}
\label{Fig:AveragePhotonNumber}
\end{figure}

\begin{figure}[h]
\includegraphics[width=8cm]{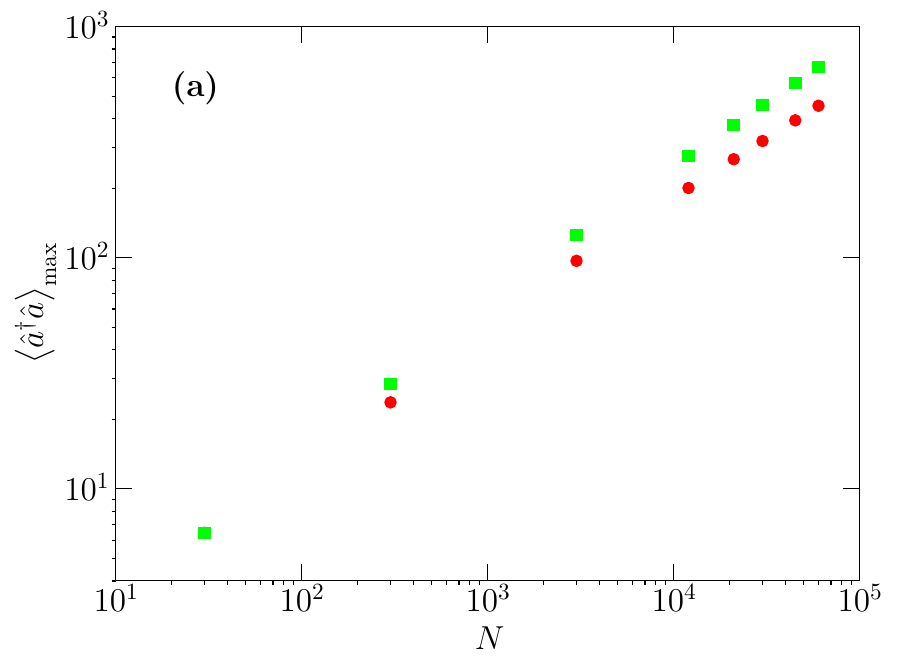}
\includegraphics[width=8cm]{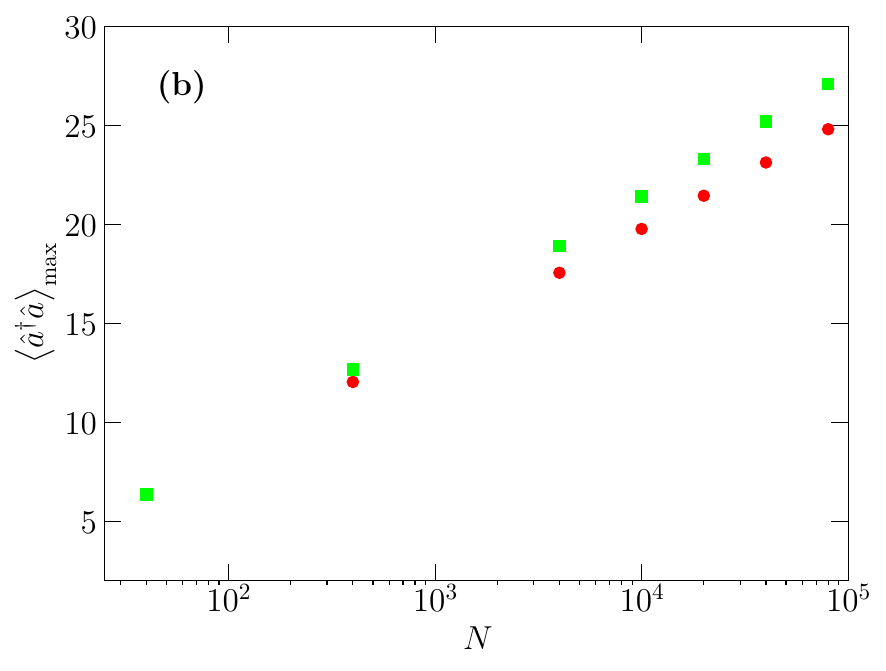}
\caption{Maximum average photon number $\left\langle \hat{a}^{\dagger} \hat{a} \right\rangle_{\rm max}$ (as seen in Fig.~\ref{Fig:AveragePhotonNumber}) for the state $\hat{U}_n(r)\ket{0}$ as a function of the maximum photon number included in our truncated Hamiltonian $\hat{H}_n$. For $n=3$, the maximum occurs at $r=0.86$, while for $n=4$, the maximum occurs at $r=0.37$. Panels (a) and (b) correspond, respectively, to $n=3$ and $n=4$. The red circles and green squares correspond, respectively, to the cases of hard and soft cutoff. Note that the $y$-axis scale is logarithmic in (a) and linear in (b). A linear fit to each data set gives $\left\langle \hat{a}^{\dagger} \hat{a} \right\rangle_{\rm max} \propto N^{0.562}$ for the red circles and $N^{0.610}$ for the green squares in Panel (a), and $\left\langle \hat{a}^{\dagger} \hat{a} \right\rangle_{\rm max} \propto \log [N]$ for both data sets in Panel (b)}
\label{Fig:AveragePhotonNumberVsTruncationSize}
\end{figure}

First, we performed numerical simulations following the method described in Sec.~\ref{Sec:Background} for the exactly solvable cases of displacement ($n=1$) and two-photon squeezing ($n=2$). Our results consistently agreed with the known solutions for these two cases. For example, when we set $r=2$ and $N=n\times 10^4$, the simulations gave average photon numbers that deviate from the exact values by $7\times 10^{-14}$ and $6\times 10^{-4}$ for the cases $n=1$ and $n=2$, respectively. This excellent agreement between the simulations and exact results serves as an additional check on the validity of the simulations with the typical parameters that we use. We will show only a few further results for the cases $n=1$ and $n=2$ below, mainly for the purpose of comparing with the case $n\geq 3$.

Next we performed simulations for the case of trisqueezing, i.e.~$n=3$. In Fig.~\ref{Fig:QW3}, we plot the Q and Wigner functions for the trisqueezed vacuum state, i.e.~$\hat{U}_3(r)\ket{0}$, for six representative values of the squeezing parameter $r$. The Q and Wigner functions serve complementary purposes. The Q function shows most clearly the coherent states that participate in a quantum superposition of such states. Note that while the appearance of the Q function suggests that it contains only probability information, i.e.~with no phase information about superpositions of coherent states, it does in fact contain (in its detailed and often not-easily-discernible features) the full information about the quantum state. On the other hand, the Wigner function tends to be nonzero in a larger area of phase space, but it conveys the presence of non-Gaussian features prominently in the form of negative values of the function.

For small values of the squeezing parameter $r$ ($r=0.1$ and 0.5), the Q and Wigner functions have a central peak at the origin of phase space with three small arms extending along three directions that make 120 degree angles with each other. This behaviour is what one would intuitively expect by considering the well-known two-photon squeezing behaviour along with the fact that $n$-th order squeezed vacuum states must have $n$-fold rotation symmetry, i.e.~they must remain unchanged when rotated by an angle of $2\pi/n$. In other words, $\hat{R}^{\dagger}(2\pi/n) \hat{H}_n \hat{R}(2\pi/2n)= \hat{H}_n$, where $\hat{R}(x)= \exp\left\{ i x \hat{a}^{\dagger} \hat{a} \right\}$, as explained in Ref.~\cite{Braunstein1987}. In fact, similar Q functions appear in Refs.~\cite{Braunstein1987,Braunstein1990}. The Wigner function takes negative values in the regions adjacent to the main peak between the directions of the three arms. Another feature that we can see in the Q function at $r=0.5$ are three broad and faint peaks along the directions between the three main arms. As we increase the squeezing parameter further ($r=0.7$), the central peak is greatly suppressed, especially as seen in the Q function, and we instead have a pattern that looks like that of a superposition of three coherent states, somewhat similar to the one synthesized in Ref.~\cite{Hofheinz}. As we increase $r$ further, the three peaks do not continue to move farther from the origin. Instead, their height decreases, and the three peaks that were faint for small $r$ increase in height. At $r=0.86$, the six peaks seem to have almost exactly the same height and shape. As we increase $r$ further ($r=1$), the three peaks that were faint become the dominant ones, while the three peaks that were dominant become faint. As we increase $r$ further ($r=1.5$), the three peaks approach the origin and merge into a single peak. At $r=1.75$ (not shown in Fig.~\ref{Fig:QW3}) the state returns almost completely to the initial state, i.e.~the vacuum state.

The periodic behaviour can also be seen clearly if we plot the occupation probabilities of the different photon number states as functions of $r$, as shown in Fig.~\ref{Fig:ProbVsSqueezeingR3}. While the oscillations are not exactly symmetric in $r$, the occupation probability of the vacuum state goes back very close to its initial value at some point. Specifically, at $r=1.75$ the probability is 0.991. On the other hand, the occupation probability of the vacuum state reaches its lowest value (below $5\times 10^{-4})$ at 0.9, which is slightly shifted from the point ($r=0.86$) where the Q and Wigner functions are closest to having six-fold symmetry. Note that in Fig.~\ref{Fig:ProbVsSqueezeingR3} the highest probability Fock state alternates between being $\ket{0}$ and $\ket{3}$. For comparison, the probability distribution for ($n=1$) keeps moving to higher photon numbers indefinitely, while for $n=2$ the probability is a monotonically decreasing function (considering only even photon numbers) and the highest probability Fock state is always $\ket{0}$.

As a check on possible finite-size effects, in Fig.~\ref{Fig:ProbVsPhotonNum3} we plot the photon number probability distribution at $r=1$ for different truncated matrices $\hat{H}_3$. We find that the majority of the data points show very weak dependence on $N$ and on whether we use a hard or soft cutoff. For all data points, the results for $N=3\times 10^4$ and $N=6\times 10^4$ are very close to each other, which suggests that these results have almost completely converged to the asymptotic values. It is worth mentioning here that we will find finite-size effects in other quantities below.

Next we turn to the case of four-photon squeezing, i.e.~$n=4$. The Q and Wigner functions are shown in Fig.~\ref{Fig:QW4}. Similarly to what we found in the case of trisqueezing, the Q function develops four small arms along the $x$ and $p$ axes in phase space. For larger values of $r$, the Q function resembles a superposition of four coherent states centered at symmetrically located points on the $x$ and $p$ axes. These peaks then gradually fade and are replaced by four peaks located symmetrically along the diagonal lines between the $x$ and $p$ axes. The Q and Wigner functions then evolve back to the vacuum state. From the plot of the occupation probabilities as functions of $r$, shown in Fig.~\ref{Fig:ProbVsSqueezeingR4}, we find that the maximum vacuum state probability at the end of the first oscillation is $0.9989$, and it occurs at $r=0.75$. It is interesting that now the occupation probabilities with $N=400$ and $N=4\times 10^4$ (with different types of cutoff) are almost indistinguishable.

The average photon number $\left\langle \hat{a}^{\dagger} \hat{a} \right\rangle$ for the state $\hat{U}_n(r)\ket{0}$, plotted as a function of the squeezing parameter $r$ in Fig.~{\ref{Fig:AveragePhotonNumber}, exhibits a few interesting features. For comparison, we include in Fig.~\ref{Fig:AveragePhotonNumber} simulation results for $n=1$ (displacement) and $n=2$ (two-photon squeezing). In these two cases, the average photon number grows indefinitely. For trisqueezing ($n=3$; shown in Fig.~7b), the average photon number exhibits irregular oscillations. A remarkable feature in the case $n=3$ is the fact that, although the height of the oscillations in Fig.~\ref{Fig:AveragePhotonNumber}(b) varies greatly for the different curves (which correspond to different truncation conditions), the overall pattern seems to be almost independent of the truncation details. This point is illustrated most clearly in the fact that, by coincidence, two of the curves (which have different values of $N$ and differ in whether they have a hard or soft cutoff) seem to overlap perfectly. Furthermore, the point at which $\left\langle \hat{a}^{\dagger} \hat{a} \right\rangle$ goes back close to zero, i.e.~the period of the oscillations, seems to be the same for all the curves in the figure. We will come back to this point shortly. As we progressively increase $n$ to $n=4$ and $n=5$, the oscillations become less irregular. More specifically, the oscillations become increasingly close to simple sinusoidal oscillations. Furthermore, the maximum value of $\left\langle \hat{a}^{\dagger} \hat{a} \right\rangle$ decreases with increasing $n$. The oscillation period also decreases with increasing $n$. A prominent feature in Fig.~\ref{Fig:AveragePhotonNumber} relates to the height of the oscillations. The maximum height for $n=3$ reaches hundreds of photons. At first sight, this value seems to contradict the results plotted in Figs.~\ref{Fig:QW3}-\ref{Fig:ProbVsPhotonNum3}, which indicate that the probability distribution remains largely confined to small photon numbers. However, the long tail of the probability distribution that extends to the maximum photon number in the simulation, i.e.~$N\sim 10^4$, is responsible for raising the average photon number to the order of hundreds.

To investigate this point further, in Fig.~\ref{Fig:AveragePhotonNumberVsTruncationSize} we plot the maximum value of $\left\langle \hat{a}^{\dagger} \hat{a} \right\rangle$ reached in the first oscillation as a function of matrix size $N$. For $n=3$, by fitting the data in Fig.~\ref{Fig:AveragePhotonNumberVsTruncationSize}(a) to a straight line, we find that to a good approximation $\left\langle \hat{a}^{\dagger} \hat{a} \right\rangle_{\rm max} \propto N^{0.562}$ for the case of a hard cutoff and $N^{0.610}$ for the case of a soft cutoff. These functions diverge as $N\to\infty$. However, it should be noted that the divergence is sub-linear. Most of the probability remains at small photon numbers, and the probability that goes to the maximum photon number decreases with increasing $N$. Still, the small probability in the tail of the distribution is sufficient to raise the average photon number to infinity in the limit $N\to\infty$. Although it is not shown in Fig.~\ref{Fig:AveragePhotonNumber}, we found that the maximum height of the $\left\langle \hat{a}^{\dagger} \hat{a} \right\rangle$ oscillations for $n=4$ also depends on the truncation details. This point is investigated further in Fig.~\ref{Fig:AveragePhotonNumberVsTruncationSize}(b). By noting that the data follows a straight line on a log-linear plot, we find that $\left\langle \hat{a}^{\dagger} \hat{a} \right\rangle_{\rm max} \propto \log [N]$, i.e.~weaker dependence that in the case $n=3$. For the case $n=5$, we found that the oscillation height in Fig.~\ref{Fig:AveragePhotonNumber} is the same for hard and soft cutoffs and for $N=5\times 10^3$ and $5\times 10^4$. We therefore conclude that the $n=5$ curve in Fig.~\ref{Fig:AveragePhotonNumber} does not contain any finite-size effects. It is quite interesting how the role of finite-size effects changes as we vary $n$: no noticeable effects for $n=1$ and $n=2$, then the clearest effects for $n=3$, milder effects for $n=4$ and back to no noticeable effects for $n=5$.

As mentioned in Sec.~\ref{Sec:Background}, we do not expect that a realistic near-future experiment will be able to observe a trisqueezed states with an average photon number of hundreds of photons, which is what the above analysis with large values of $N$ predicts. For example, Ref.~\cite{Eriksson} reached $r=0.13$ for $n=3$, while Ref.~\cite{Bazavan} reached $r=0.19$ for $n=3$ and $r=0.054$ for $n=4$. For $n=2$, the record squeezing parameter for light is $r=1.7$ \cite{Vahlbruch}. These numbers cast doubt on the possibility of observing any effects that involve large photon numbers. However, since residual terms in a realistic system Hamiltonian could effectively impose a cutoff of only tens or hundreds of photons, it should be possible to observe the oscillatory behaviour of the squeezing dynamics without the need to reach high photon numbers. As the exact behaviour will depend on the details of the actual physical system, we shall not perform an analysis of any specific physical system here.

\section{Discussion}
\label{Sec:Discussion}

In this section we discuss a few additional aspects of the results presented in Section \ref{Sec:Results}.

\subsection{Mathematical aspects of truncated simulations}

The periodic behaviour that we observe in our simulations is in stark contrast to the displacement ($n=1$) and two-photon squeezing ($n=2$) dynamics. In those two cases, the progression of the displacement or squeezing, as seen for example in the Q and Wigner functions, continues indefinitely. The photon number distribution keeps evolving towards ever larger numbers, asymptotically approaching infinity.

An immediate question that comes to mind in this case is whether our results are a mathematical artefact of our numerical simulation approach, especially considering the mathematical complications that were encountered in previous studies on generalized squeezing. A strong piece of evidence that our results are not mathematical artefacts is the near-perfect agreement between the simulation results obtained under greatly different conditions. As an example, one can take Fig.~\ref{Fig:ProbVsSqueezeingR4}, where two different simulations produced almost identical results, even though one simulation used 400 basis states with a soft cutoff, while the other used $4\times 10^4$ basis states with a hard cutoff. If the results were mathematical artefacts of our truncation, we would expect to see pronounced finite-size effects that are different between these two simulations of vastly different sizes. Another somewhat paradoxical situation is the fact that in Fig.~\ref{Fig:AveragePhotonNumber} the oscillation height is truncation dependent but the oscillation period is truncation independent. The latter suggests that the oscillations are not a mathematical artefact.

In this context, we also note another remarkable aspect of the fact that the average photon number grows indefinitely with increasing $N$. Normally, when a simulated quantity reaches the edge of the simulation space and comes back, it signals that the simulation results cannot be trusted after the reflection off the artificially imposed boundary of the simulation space. For example, we encounter this situation in the case $n=2$ if we keep increasing $r$. In that case, however, the $r$ value at which the state comes back close to the initial state also grows indefinitely with increasing $N$. In the case of the $n=3$ oscillations, the system consistently goes back to the vacuum state at $r=1.75$. One intuitive way to think about the divergence and subsequent decrease of the average photon number is as follows: because of the large values of $H_{k,k+n}^{(n)}$ at large values of $k$, the large-$k$ part of the Hilbert space can be thought of as a single, strongly-coupled subspace as seen from the small-$k$ subspace. As a result, although the dynamics inside the large-$k$ subspace differs greatly for different truncation conditions, the dynamics between the low and high-$k$ subspaces is almost truncation independent.

\subsection{Eigenvalues and eigenstates of squeezing generator}

\begin{figure}[h]
\includegraphics[width=9cm]{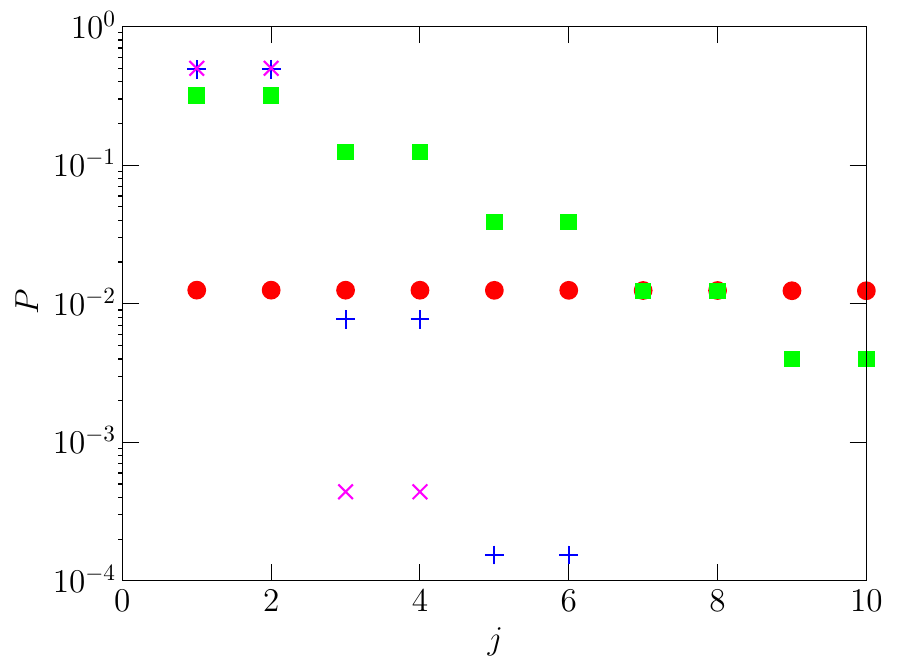}
\caption{Highest ten probabilities of the Hamiltonian eigenstates based on the overlap between these eigenstates and the vacuum state $\ket{0}$. The $x$-axis label $j$ enumerates the plotted probabilities. The red circles, green squares, blue + signs and magenta $\times$ symbols correspond, respectively, to $n=1$, 2, 3 and 4. All data sets are for $N=n\times 10^4$ with a hard cutoff. For $n=1$, the state $\ket{0}$ is a superposition of a large number of Hamiltonian eigenstates. For $n=3$ and $n=4$, the state $\ket{0}$ is, to a good approximation, a superposition of only two Hamiltonian eigenstates, a situation that leads to simple oscillations. For $n=2$, the pattern is different from both those for $n=1$ and $n\geq 3$, which makes it more difficult to make definitive statements based on this figure alone.}
\label{Fig:EigenstatProbabilities}
\end{figure}

\begin{figure}[h]
\includegraphics[width=8cm]{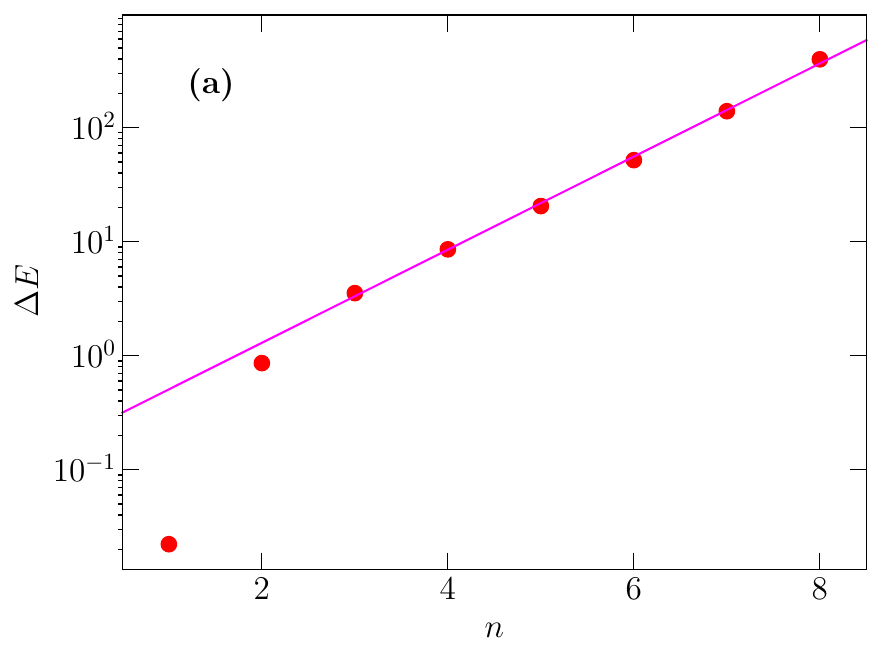}
\includegraphics[width=8cm]{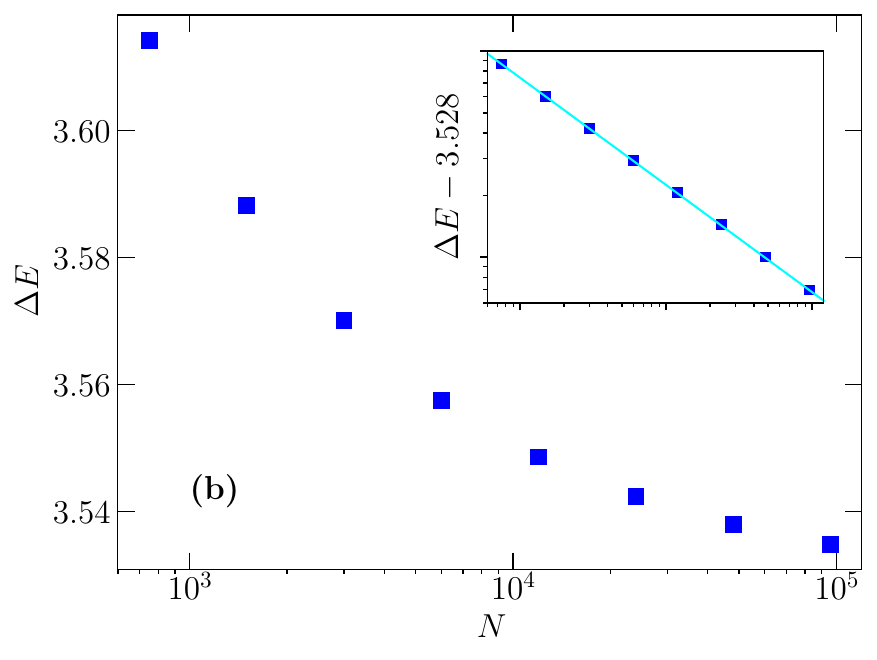}
\caption{The difference $\Delta E$ between the two eigenvalues that have the highest probabilities in the vacuum state. In Panel (a), we plot $\Delta E$ as a function of squeezing order $n$. Starting from $n=3$, the data clearly follow an exponential scaling law, as illustrated by the solid line, which was obtained by fitting the data to a straight line. The data up to $n=6$ were obtained using $N=n\times 2\times 10^4$, after which the two eigenvalues became asymmetric about zero, indicating calculation errors. Therefore, for $n=7$ and $n=8$, we used $N=n\times 2\times 10^3$. In Panel (b), we plot $\Delta E$ as a function of truncation size $N$ for $n=3$. The inset shows a log-log plot of $\Delta E - 3.528$ (which was obtained as a fitting parameter) as a function of $N$. The $x$-axis range is the same as the main plot, and the $y$-axis range is $[6\times 10^{-3},10^{-1}]$. The straight line fit in this log-log plot indicates that $\Delta E$ converges to 3.528 when $N\to\infty$. All data points in this figure were calculated with a hard cutoff.}
\label{Fig:EigenvalueDifference}
\end{figure}

\begin{figure}[h]
\includegraphics[width=7cm]{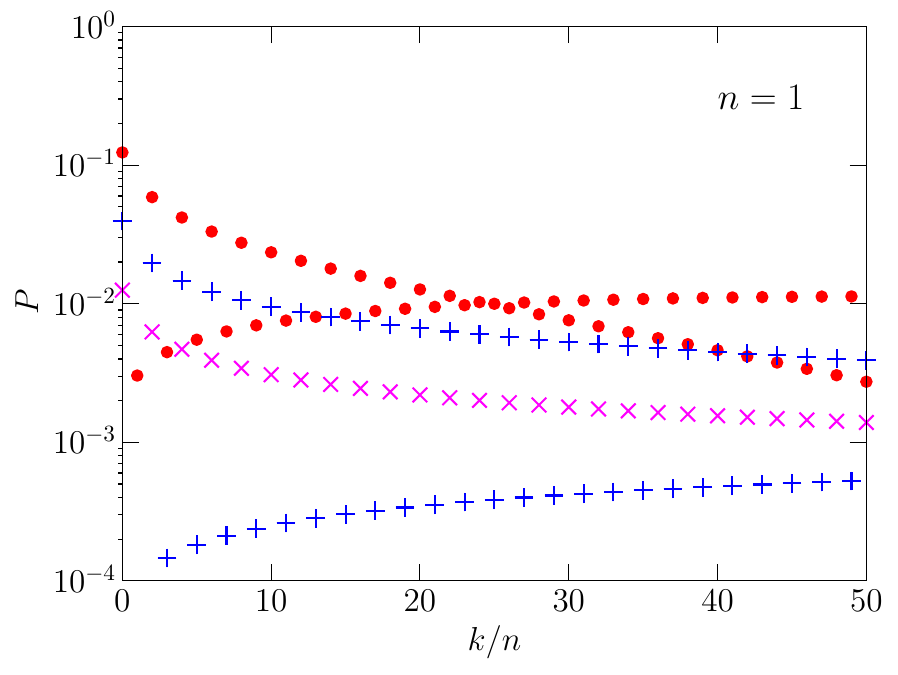}
\includegraphics[width=7cm]{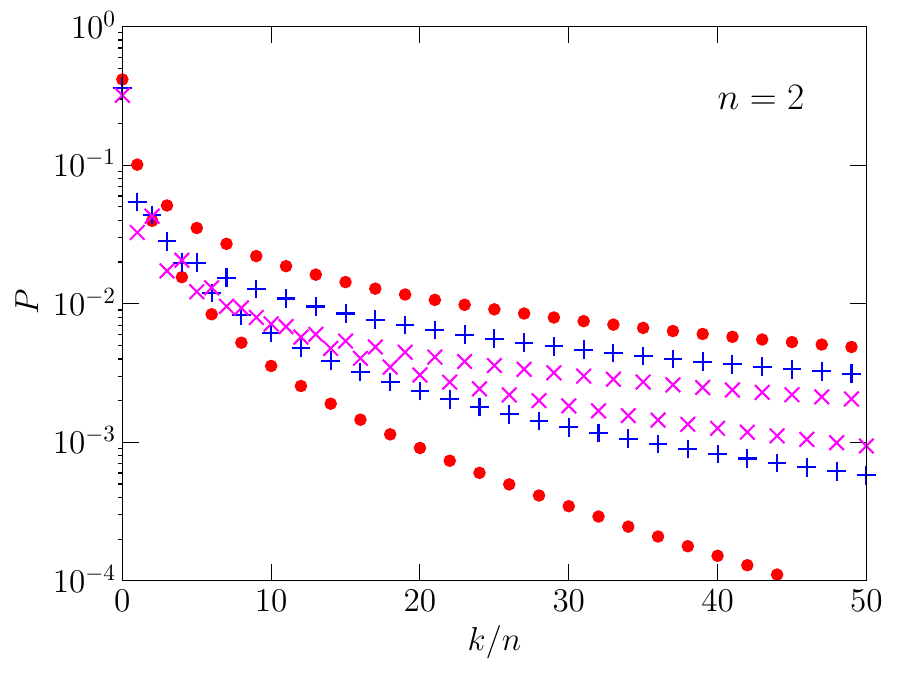}
\includegraphics[width=7cm]{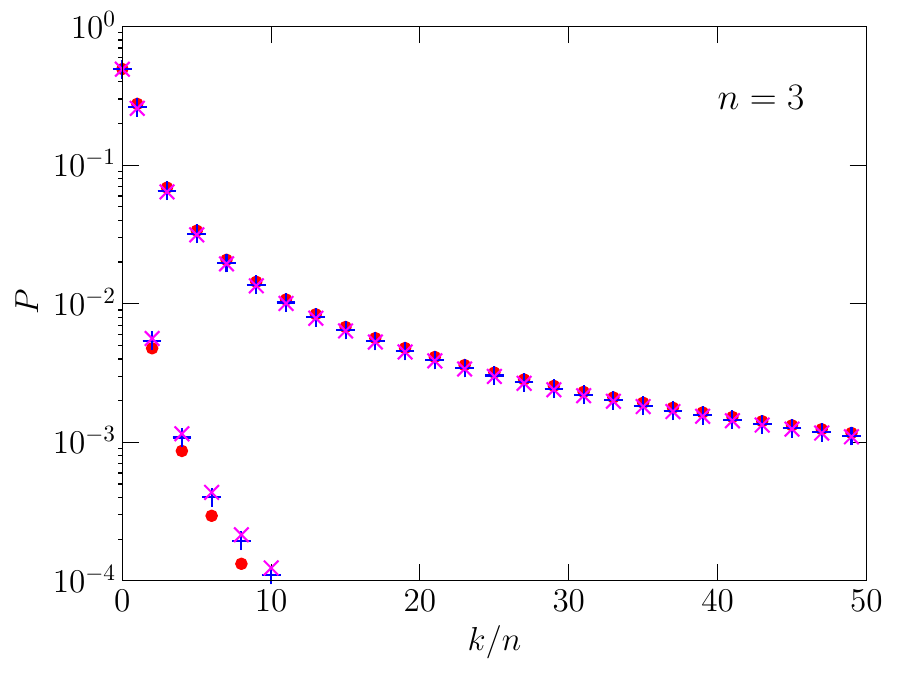}
\includegraphics[width=7cm]{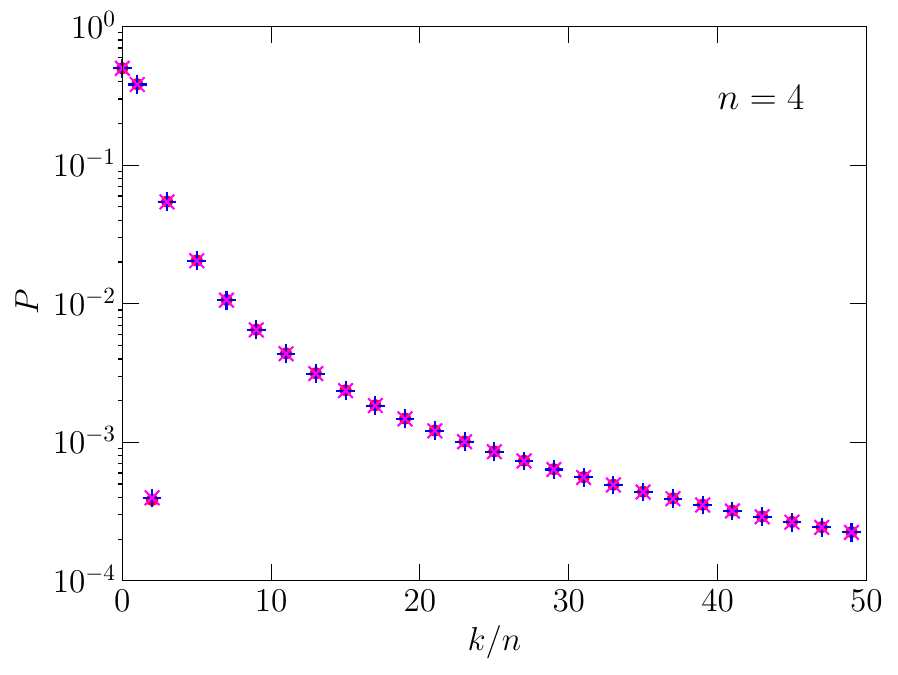}
\caption{Probability distribution (as a function of photon number $k$) of the Hamiltonian eigenstate that has the maximum overlap with the vacuum state $\ket{0}$. The red circles, blue + signs and magenta $\times$ symbols correspond, respectively, to $N=n\times 100$, $n\times 1000$ and $n\times 10^4$, all with a hard cutoff. Note that the probability distributions oscillate as functions of $k$, such that there are data points with $P<10^{-4}$ that are not shown in this figure. For $n=1$, the data does not seem to decrease asymptotically with increasing $k$. Furthermore, the probabilities decrease significantly with increasing $N$. These two facts show that the most important eigenstates are delocalized over the full truncated Hilbert space. For $n=3$ and $n=4$, the probabilities decrease quickly as functions of $k$ and exhibit only weak dependence on $N$. For $n=2$, the pattern is different from both those for $n=1$ and $n\geq 3$, which makes it more difficult to make definitive statements based on this figure alone.}
\label{Fig:MainEigenstatProbDist}
\end{figure}

To further investigate the difference between the dynamics for different values of $n$, we analyze the eigenstates of the Hamiltonian $\hat{H}_n$. Of particular interest is the overlap between the vacuum state $\ket{0}$, which was used as the initial state in all of our simulations, and the different eigenstates. In Fig.~\ref{Fig:EigenstatProbabilities}, we plot the highest ten eigenstate probabilities in the vacuum state. For $n=1$, the vacuum state is a superposition of a large number of eigenstates, while for $n\geq 3$, the vacuum state is a superposition of mainly two eigenstates. When a quantum system is in a superposition of two Hamiltonian eigenstates, it is natural that the dynamics will exhibit simple oscillations at a frequency given by the difference between the two corresponding eigenvalues. The difference ($\Delta E$) between the two most relevant eigenvalues is plotted in Fig.~\ref{Fig:EigenvalueDifference}. For $n=3$ (analyzed in detail in Fig.~\ref{Fig:EigenvalueDifference}b), we find that $\Delta E=3.528$, which gives an oscillation period of $r=2\pi/\Delta E=1.781$, close to the period observed in Figs.~\ref{Fig:ProbVsSqueezeingR3} and \ref{Fig:AveragePhotonNumber}. From the plot of the eigenvalue difference $\Delta E$ as a function of $n$ (Fig.~\ref{Fig:EigenvalueDifference}a), we find that for $n\geq 3$ the data follows an exponential scaling law, which means that the oscillation period decreases rapidly as we increase $n$.

In the same line of investigating Hamiltonian eigenstates, in Fig.~\ref{Fig:MainEigenstatProbDist} we plot the probability distribution of the highest-probability eigenstate as a function of photon number $k$. The contrast between the cases $n=1$ and $n\geq 3$ is clear. In the former case, the probability distribution extends to large values of $k$, while in the latter case, the probability distribution is strongly localized at small values of $k$. The case $n=2$ lies somewhere in between the extremes of the cases $n=1$ and $n\geq 3$, which makes it difficult to make statements about the case $n=2$ from this eigenstate analysis. Considering the two most relevant eigenstates for $n=3$, they are complex conjugates of each other, and each has $\left\langle \hat{a}^{\dagger} \hat{a} \right\rangle = 112.6$. If we approximate the initial state as a superposition of only these two states, we expect that $\left\langle \hat{a}^{\dagger} \hat{a} \right\rangle$ will exhibit simple sinusoidal oscillations with amplitude 225. The maximum amplitude for the corresponding curve in Fig.~\ref{Fig:AveragePhotonNumber} is 320. In other words, the two main eigenstates account for 70\% of the oscillation amplitude. Indeed, the irregularity in the oscillation pattern is a clear indication that more than two Hamiltonian eigenstates make substantial contributions. For $n=4$, the main two eigenstates have $\left\langle \hat{a}^{\dagger} \hat{a} \right\rangle = 10.6$ and hence account for 91\% of the oscillation amplitude. For $n=5$, the main two eigenstates have $\left\langle \hat{a}^{\dagger} \hat{a} \right\rangle = 4.26$ and account for 99\% of the oscillation amplitude. We also note here that $\left\langle \hat{a}^{\dagger} \hat{a} \right\rangle$ for the two main eigenstates increases with $N$ following similar scaling laws as those explained in Fig.~\ref{Fig:AveragePhotonNumberVsTruncationSize}.

\subsection{Analogy with wave propagation}

One result that we did not show in detail but can be inferred from Fig.~\ref{Fig:ProbVsPhotonNum3} is that the probability distribution as a function of photon number $k$ decreases monotonically for small values of $r$ but develops an oscillatory pattern as $r$ is increased. In fact, the oscillations in the long tail of the probability distribution seem to be present even at small values of $r$, but they become noticeable only for larger values of $r$. For example, for $n=3$, the oscillations reach the states with photon number below 30 only for $r>0.2$. This oscillatory pattern in the probability distribution that appears above a certain value of $r$ suggests an interference effect between waves propagating in opposite directions. More specifically, it suggests that the wave of probability amplitudes initially propagates from low to high values of $k$ but is reflected at some point and the interference between the forward and backward propagating waves creates the oscillations. Note that this situation does not occur for displacement ($n=1$) and two-photon squeezing ($n=2$). This situation also raises the question of where the reflection happens. Importantly, if the reflection happens at the end of the photon number ladder in our truncated Hamiltonian, it could be indicative of finite-size effects, and we have to treat the results with caution. If the reflection point is some naturally occurring photon number (which can depend on $n$ but does not depend on $N$), it would be an indication that this reflection is a physical result as opposed to a mathematical artefact. The fact that the probability pattern depends only weakly on $N$, which covers three orders of magnitude in our simulations, as shown in Fig.~\ref{Fig:ProbVsSqueezeingR4}, leads us to believe that the reflection is a physical phenomenon rather than being a finite-size effect.

Indeed, there could be an analogy with the problem of Bloch oscillations, in which a wave packet propagating in a nonuniform lattice cannot sustain its forward propagation if it cannot adapt to the changing band structure, and instead it gets reflected and starts propagating in the opposite direction.

\subsection{Classical analogy}

In the context of the Bloch oscillation analogy, it is interesting to consider the following classical approximation: the matrix elements in the Hamiltonian are given by Eq.~(\ref{Eq:MatrixElements}), which asymptotically approaches
\begin{equation}
\lim_{k\to\infty} H_{k,k+n}^{(n)} = k^{n/2}.
\end{equation}
If we consider an analogy of a particle propagating in a lattice or a spin wave propagating in a spin chain with nearest-neighbour matrix elements all (approximately) given by $J$ in a certain region in space, the maximum group velocity of propagation is given by $2J$ \cite{Ashhab2012}. If we now consider a classical approximation where the velocity $v$ is given by
\begin{equation}
v = \frac{dx}{dt} = \alpha x^{n/2},
\end{equation}
we find that the position $x(t)$ at time $t$ is given by
\begin{equation}
x(t) = \left\{
\begin{array}{ll}
\left( \sqrt{x(0)} + 2 \alpha t \right)^2, & n=1, \\
x(0) e^{\alpha t}, & n=2, \\
\left[ x^{-(n/2-1)}(0) - \left( \frac{n}{2} - 1 \right) \alpha t \right]^{-1/(n/2-1)}, & n\geq 3.
\end{array}
\right.
\end{equation}
The function $x(t)$ for $n=1$ exhibits the correct quadratic growth of the position (which represents the number of photons) with time, while the functions for $n\geq 2$ do not accurately describe the quantum dynamics. It is nevertheless interesting to note that in the cases $n=1$ and $n=2$ the position grows indefinitely up to $t\to\infty$, while for $n\geq 3$, it diverges at a finite value of $t$ (assuming that $x(0)\neq 0$ for $n\geq 2$), indicating a qualitative difference between the cases $n\leq 2$ and $n\geq 3$.

\subsection{Squeezing using the operator $\hat{p}^n$}

\begin{figure}[h]
\includegraphics[width=5.5cm]{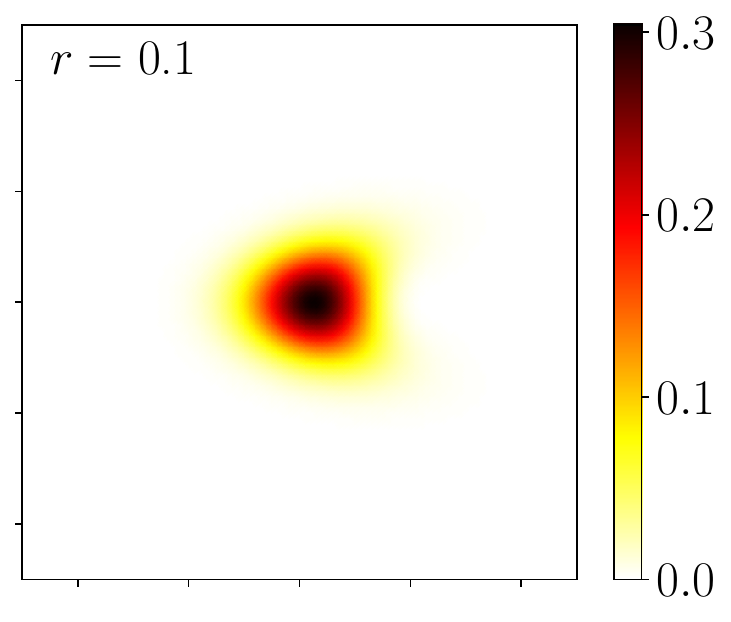}
\includegraphics[width=5.5cm]{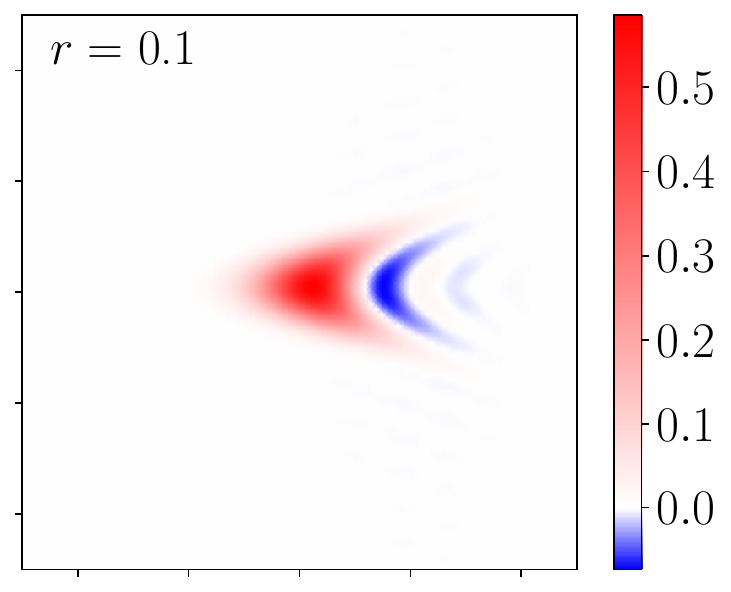}
\includegraphics[width=5.5cm]{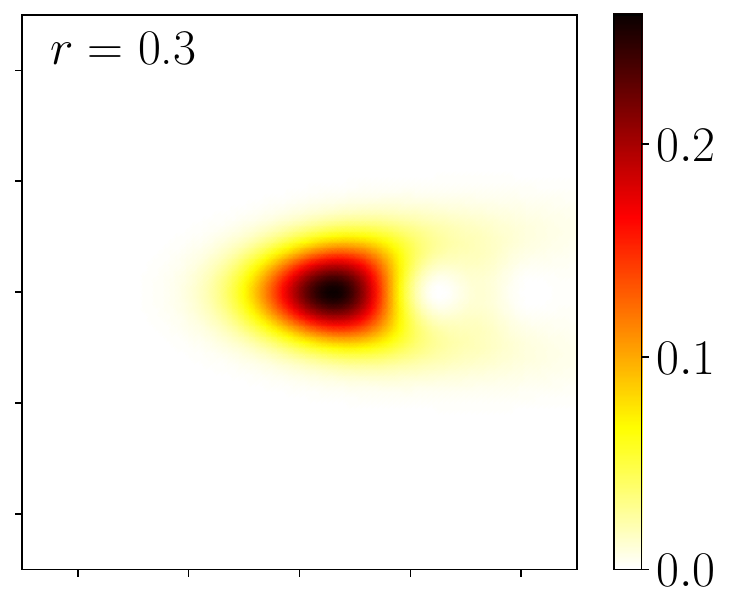}
\includegraphics[width=5.5cm]{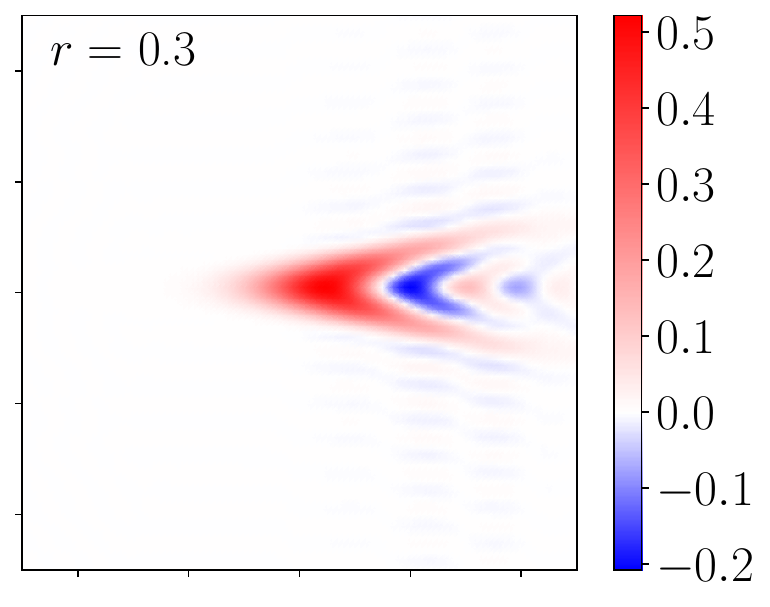}
\includegraphics[width=5.5cm]{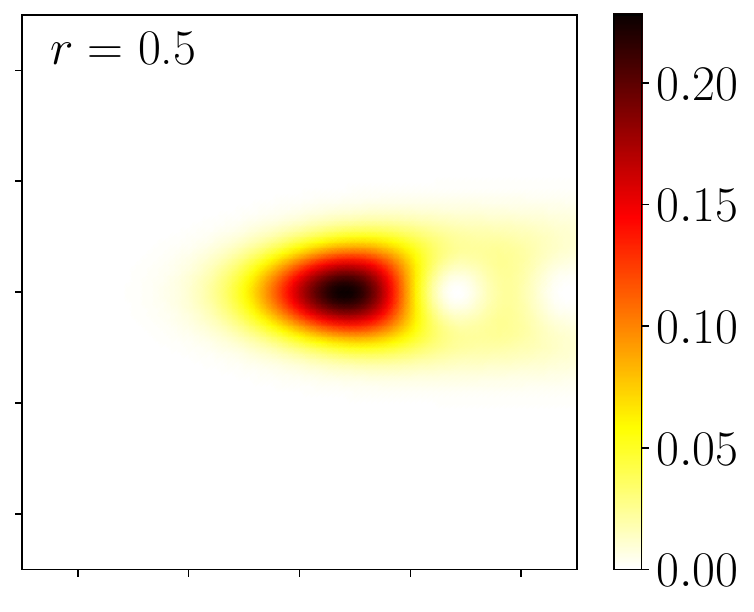}
\includegraphics[width=5.5cm]{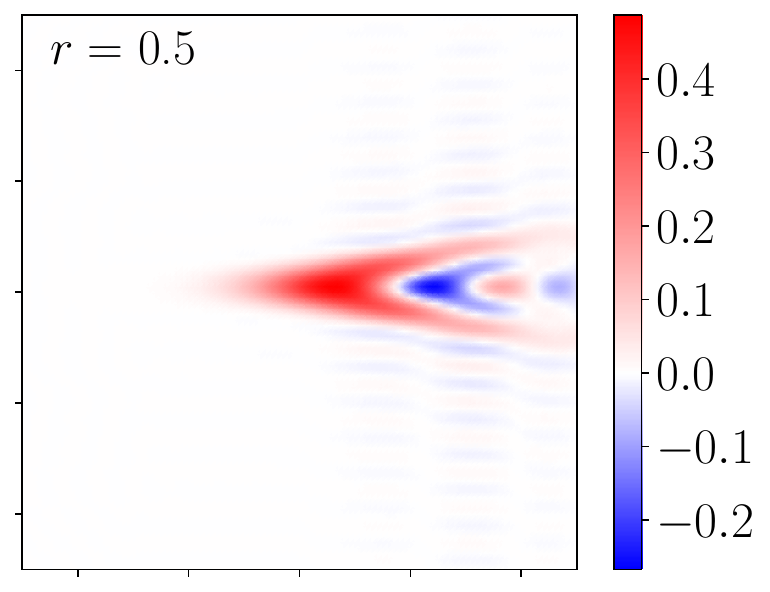}
\caption{Q and Wigner functions for the squeezed vacuum state obtained using the squeezing operator in Eq.~(\ref{Eq:Hamiltonian_n_Alternative}) with $n=3$, also known as cubic phase state, for $r=0.1$, 0.3 and 0.5. Unlike the trisqueezed state discussed in Sec.~\ref{Sec:Results}, the cubic phase state does not have three-fold symmetry, and it does not undergo periodic dynamics.}
\label{Fig:QWX3}
\end{figure}

It is interesting to consider the alternative effective Hamiltonian \cite{Gottesman}
\begin{equation}
\hat{H}_n = \left(\frac{\hat{a}^\dagger - \hat{a}}{i} \right)^n.
\label{Eq:Hamiltonian_n_Alternative}
\end{equation}
Since this operator can be expressed as $\hat{p}^n$, where $\hat{p}$ is a rescaled momentum operator, it is possible to perform analytical derivations in this case, e.g.~using wave functions in momentum space. One can, with some algebra, show that the average photon number in the squeezed vacuum state is given by
\begin{equation}
\left\langle \hat{a}^{\dagger} \hat{a} \right\rangle = (2n-3)!! n^2 r^2,
\label{Eq:AveragePhotonNumberPnSqueezing}
\end{equation}
where $m!!=m(m-2)\cdots 1$. In other words, the squeezed state keeps growing and does not undergo periodic dynamics. It should be noted, however, that these squeezed states are qualitatively different from those generated by the squeezing operator in Eq.~(\ref{Eq:Hamiltonian_n}). The operator in Eq.~(\ref{Eq:Hamiltonian_n_Alternative}) contains terms that create or annihilate different numbers of photons. For example, for $n=3$, there are single-photon terms. As a result, the squeezed states do not have three-fold rotation symmetry anymore, as can be seen in Fig.~\ref{Fig:QWX3}. Similar states were reported in Ref.~\cite{Eriksson}. It is interesting to note here that the average photon number in Eq.~(\ref{Eq:AveragePhotonNumberPnSqueezing}) always grows quadratically in $r$, while $n$ determines the prefactor in the formula.

It should also be noted that the operator in Eq.~(\ref{Eq:Hamiltonian_n}) arises more naturally in experiment than the operator in Eq.~(\ref{Eq:Hamiltonian_n_Alternative}). The effective Hamiltonian that generates the squeezing is typically derived from the Hamiltonian
\begin{equation}
\hat{H} = \hbar \omega \hat{a}^{\dagger} \hat{a} + \Omega \left(\frac{\hat{a}^\dagger - \hat{a}}{i} \right)^n \sin n\omega t.
\end{equation}
Upon making the rotating-wave approximation, the only terms that are time independent in the rotating frame will be ones containing $(\hat{a}^{\dagger})^n$ and $\hat{a}^n$, leading to the effective Hamiltonian in Eq.~(\ref{Eq:Hamiltonian_n}).

\subsection{Avoiding the oscillations with generalized squeezing}

\begin{figure}[h]
\includegraphics[width=5.5cm]{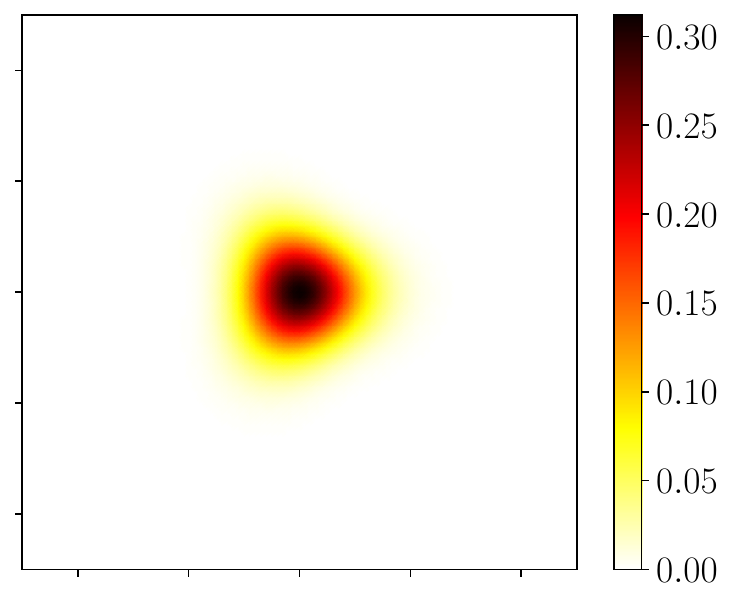}
\includegraphics[width=5.5cm]{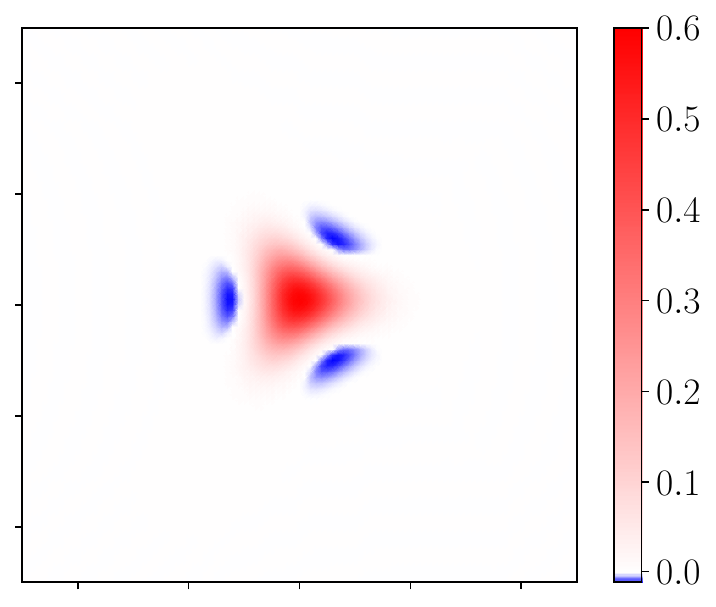}
\includegraphics[width=5.5cm]{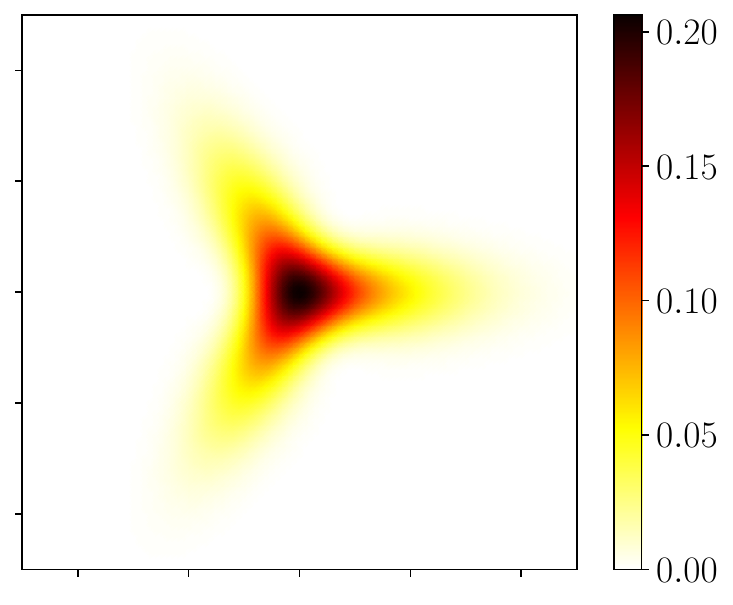}
\includegraphics[width=5.5cm]{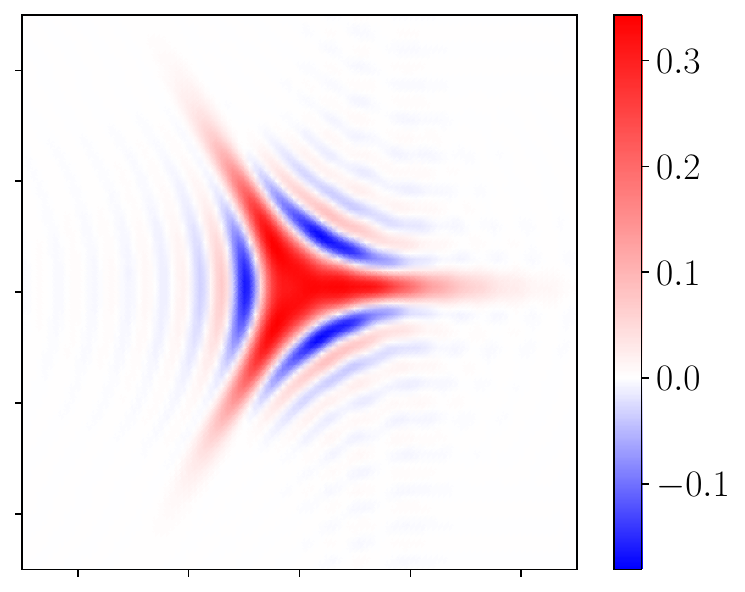}
\includegraphics[width=5.5cm]{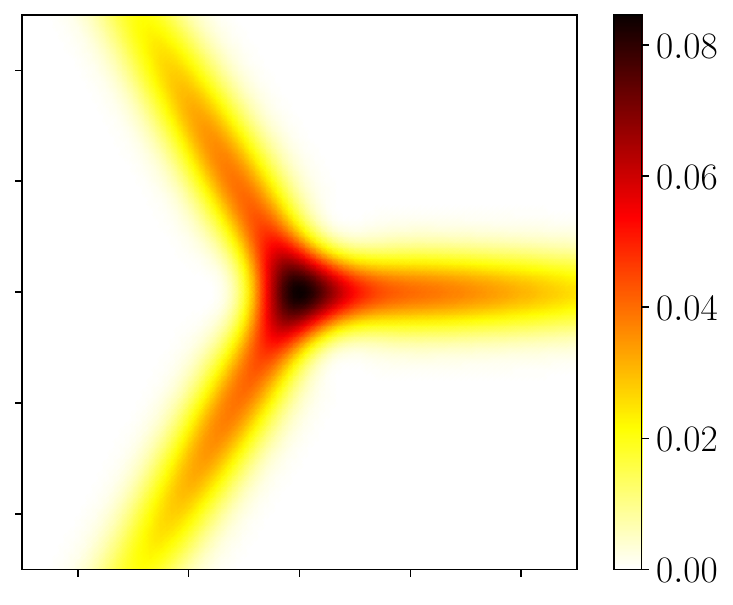}
\includegraphics[width=5.5cm]{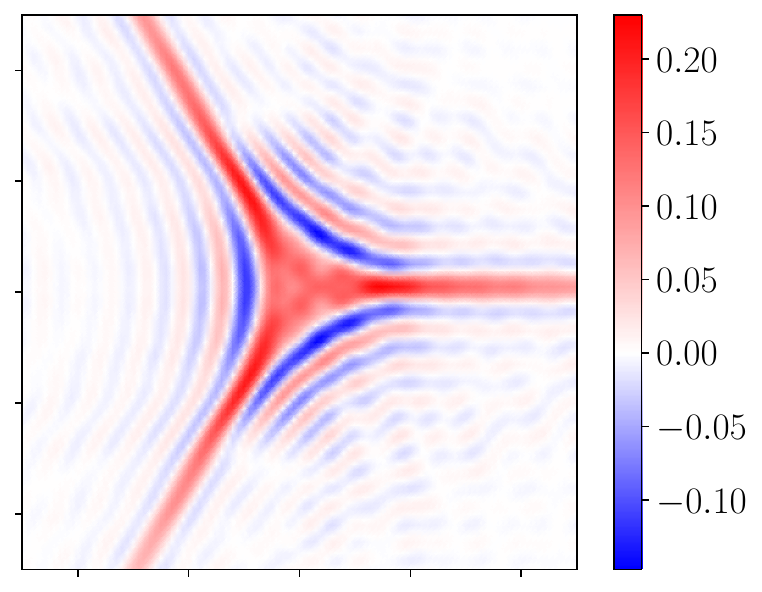}
\caption{Q and Wigner functions for quantum states that are constructed by taking the amplitudes of the states $\ket{0}$, $\ket{2}$, $\ket{4}$ ... in the state $\hat{U}_2(r)\ket{0}$ and assigning these amplitudes to the states $\ket{0}$, $\ket{3}$, $\ket{6}$ ... In other words, these are artificially constructed trisqueezed states. The squeezing parameters used to generate these plots are given by $r=0.1$ (top row), 0.5 (middle row) and 1.0 (bottom row).}
\label{Fig:QW3from2}
\end{figure}

We finally note that one can generate squeezed states whose Q and Wigner functions keep getting elongated indefinitely, even for $n\geq 3$, by using different Hamiltonians. For example, we can take inspiration from the dynamics of two-photon squeezing and design a Hamiltonian whose matrix elements vanish except for $H_{3k,3(k+1)}=\sqrt{(2k+1)(2k+2)}$ for all integer values of $k$. If we now take this Hamiltonian and the initial state $\ket{0}$, we find that the probability amplitudes of the states $\ket{0},\ket{3},\ket{6},...$ grow in exactly the same way that the amplitudes of the states $\ket{0},\ket{2},\ket{4},...$ grow with two-photon squeezing. As a result, the three arms of the trisqueezed state keep growing indefinitely. An example of such states is shown in Fig.~\ref{Fig:QW3from2}.

Another way to look at the possibility of designing indefinitely growing generalized squeezed states is as a case of reverse engineering: if we know the quantum state evolution that we want to induce, we can differentiate the state and substitute it in the Schr\"odinger equation to obtain the necessary Hamiltonian. Our results demonstrate that the intuitive, straightforward generalization of displacement and squeezing protocols cannot be used to obtained arbitrarily squeezed states with $n\geq 3$. New protocols must be developed for this purpose. For near-future applications, e.g.~for photon numbers below 100, optimal control theory (OCT) can be a powerful method to design control protocols that achieve the desired goal. For example, OCT was used to generate arbitrary few-photon states in a superconducting resonator controlled via a qubit \cite{Heeres}.

\section{Conclusion}
\label{Sec:Conclusion}

We have demonstrated, using a variety of numerical simulations and additional numerical checks, that generalized squeezing with $n\geq 3$ is qualitatively different from the lower-order operations of displacement and two-photon squeezing. The indefinite growth of the photon number for the low-order operations is replaced with oscillatory dynamics for higher-order operations. Although some quantities exhibit finite-size effects and would extrapolate to infinity for an infinite simulation size, several features in the results point to the oscillations being a physical phenomenon that is mathematically well behaved and should be observable in future experiments. As experimental capabilities on nonlinear quantum optics are advancing rapidly, we expect that our results will serve as an important reference point for the development of generalized squeezing protocols in the future.

\section*{Acknowledgment}

We would like to thank Daniel Braak, Suguru Endo, Lin Tian and Fumiki Yoshihara for useful discussions. This work was supported by Japan's Ministry of Education, Culture, Sports, Science and Technology (MEXT) Quantum Leap Flagship Program Grant Number JPMXS0120319794.

\section*{Data availability}

The datasets generated and/or analysed during the current study are available from the corresponding author on reasonable request.

\end{document}